\documentclass[twocolumn,10pt]{wlscirep}
\usepackage{bm}
\usepackage{subcaption}
\usepackage{graphicx}
\usepackage{epstopdf}
\usepackage{wrapfig}
\usepackage{array} 
\usepackage{listings}
\usepackage{csquotes}
\usepackage[para,online,flushleft]{threeparttablex}
\usepackage{booktabs,dcolumn}
\usepackage{color}
\usepackage{textpos}
\usepackage{multirow,bigdelim}
\usepackage{float}
\usepackage{upgreek}
\usepackage{xcolor}
\usepackage{mathrsfs}
\usepackage{slashed}
\usepackage{slashbox}
\usepackage{caption}
\usepackage{algorithm}
\usepackage{algorithmicx}
\usepackage{algpseudocode}
\usepackage{amsmath}     
\usepackage{multicol}
\usepackage{orcidlink}
\usepackage{adjustbox}

\usepackage{hyperref}
\hypersetup{breaklinks=true,colorlinks=true,linkcolor=blue,citecolor=blue,filecolor=magenta,urlcolor=blue}
\title{%Probing single-$\Lambda$ hypernuclei via neural-network quantum states
Machine learning the single-$\Lambda$ hypernuclei with neural-network quantum states}

\author[1]{Zi-Xiao Zhang~\orcidlink{0009-0003-4773-1505}}
\author[2]{Yi-Long Yang~\orcidlink{0000-0002-5065-1309}}
\author[1,3 *]{Wan-Bing He~\orcidlink{0000-0002-3854-4965}}
\author[2]{Peng-Wei Zhao~\orcidlink{0000-0001-8243-2381}}
\author[4]{Bing-Nan Lu~\orcidlink{0000-0001-7890-4948  }}
\author[1,3,$\dagger$]{Yu-Gang Ma~\orcidlink{0000-0002-0233-9900}}

\affil[1]{Key Laboratory of Nuclear Physics and Ion-beam Application (MOE), Institute of Modern Physics, Fudan University, Shanghai 200433, China}
\affil[2]{State Key Laboratory of Nuclear Physics and Technology, School of Physics, Peking University, Beijing 100871, China}
\affil[3]{Shanghai Research Center for Theoretical Nuclear Physics, NSFC and Fudan University, Shanghai 200438, China}
\affil[4]{Graduate School of China Academy of Engineering Physics, Beijing 100193, China}

\affil[*]{hewanbing@fudan.edu.cn}
\affil[$\dagger$]{mayugang@fudan.edu.cn}

\begin{abstract}
 Single-$\Lambda$ hypernuclei are the most straightforward extension of atomic nuclei. A thorough description of baryonic system beyond first-generation quark sector is indispensable for the maturation of nuclear $ab$ $initio$ methods. This study pioneers the application of neural-network quantum states to hypernuclei, with trainable parameters determined by variational Monte Carlo approach (VMC-NQS). In order to reduce the numerical uncertainty and treat the nucleons and hyperons in a unified manner, spinor 
 grouping (SG) method is proposed to analytically integrate out isospin degrees of freedom. A novel spin purification scheme is developed to address the severe spin contamination occurring in standard energy minimization due to the weakly bound characteristic of light single-$\Lambda$ hypernuclei. The $\Lambda$ separation energies of $s$-shell hypernuclei are computed with one-thousandth level accuracy and benchmarked against existing results from stochastic variational method, showing superior performance. By comparing two different sets of Hamiltonian based on pionless effective field theory ($\slashed{\pi}$EFT), we choose an optimal model and further carry out calculations of selected $p$-shell charge-symmetric hypernuclei with mass number up to 13, exhibiting satisfactory consistency with experimental results. Our findings underscore the potential of VMC-NQS family in approaching exact solution of few-body systems and the accuracy of $\slashed{\pi}$EFT in modeling hypernuclei. This is crucial for understanding hyperon-nucleon-nucleon and hyperon-hyperon-nucleon interactions, providing a powerful tool for precisely predicting the properties of multi-strangeness hypernuclei.
    
\end{abstract}

\begin{document}

\flushbottom
\maketitle
\thispagestyle{empty}

\section*{Introduction}
As self-bound systems composed of nucleons and the lightest hyperons, $\Lambda$ hypernuclei introduce a new dimension to the nuclear chart, extending the research of strong interaction beyond the realm of up and down quarks\cite{RevModPhys.88.035004}. Serving as an ideal test ground for hyperon-nucleon ($YN$) interactions and nuclear many-body techniques, the study of light single-$\Lambda$ hypernuclei has emerged as a unique frontier in nuclear physics\cite{Yu-Gang2017, Chen:2023mel, PhysRevLett.134.022301}. Thus far, relativistic heavy-ion collision experiments have verified the generation of light single-$\Lambda$ hypernuclei (A$\leq$4) in the hot hadronic matter created at the Relativistic Heavy Ion Collider (RHIC) \cite{STAR2010, STAR2024} and the Large
Hadron Collider (LHC) \cite{PhysRevLett.134.162301}. A deepened understanding upon light hypernuclei is crucial for elucidating the impact of strangeness in early universe evolution\cite{Chen:2018tnh, Chen:2024aom}. Moreover, compared with nuclear force, the $YN$ interactions are relatively weaker, leading to halo structures in $A\leq7$ hypernuclei \cite{2025halo}, which are of great interest as they further shed light on baryon-baryon interactions and production mechanisms of light clusters in heavy-ion collisions \cite{Yu-GangMa2023, LIU2024138855, PhysRevLett.134.022301, Chen2025}.

Driven by the remarkable success of $\Lambda N$ interaction constructed based on phenomenological description and effective field theory, the past two decades have witnessed an enduring interest and significant progresses in probing $s$- and $p$-shell single-$\Lambda$ hypernuclei.
Faddeev ($^3_\Lambda$H) \cite{Meoto_2020} and Faddeev-Yakubovsky ($^4_\Lambda$H and $^4_\Lambda$He) \cite{Faddev} calculations have been performed using Ge'fand-Levitan-Marchenko theory and Nijmegen SC $YN$ interactions.
$^{10}_\Lambda$B and $^{10}_\Lambda$Be have also been explored based on the cluster model \cite{Hiyama_2012}. In addition, quantum Monte Carlo framework, including variational Monte Carlo (VMC) \cite{2024vmc, Usmani_2006, Usmani_2008} and auxiliary-field diffusion Monte Carlo (AFDMC) \cite{LONARDONI2013243, PhysRevC89}, has been attempted based on baryonic correlation operators to address single-$\Lambda$ hypernuclei. Recently, significant efforts have been made to probe hypernuclei within nuclear lattice effective field theory (NLEFT) \cite{Frame2020ImpurityLM, Hildenbrand2022, Hildenbrand2024, tong2025abinitiocalculationhyperneutron}, Gamow shell model (GSM) \cite{LI2025139708}, and the no-core shell model (NCSM) \cite{Le2025, PRLLe}. Among the variety of theoretical frameworks, the stochastic variational method (SVM) has been widely used as an ideal toolbox to tackle light $\Lambda$ hypernuclei \cite{Nemura1999StudyOL, PhysRevC.106.L031001, Contessi2019csf, Schafer2020rba, Schfer2021ConsequencesOI}, successfully resolving the long-standing $^5_\Lambda$He overbinding problem based on pionless effective field theory ($\slashed{\pi}$EFT) \cite{SVMPRL}. SVM is achieved by optimizing the ansatz, given as a combination of correlated Gaussian bases, through a gambling procedure. However, the priori restriction that the virtual correlated motion of baryons can be represented by correlated Gaussian bases inherently constrains its ability to describe the finer details of baryonic nodal surface. Hence, an unprejudiced representation of wavefunction is highly desired for variational calculations.

The generality and flexibility of neural-networks offers a compact yet unbiased characterization of many-body quantum eigenstates into complex probability amplitudes\cite{neuralsci}, neural-network quantum states (NQS), which makes it a suitable alternative for the ansatz in VMC calculation, dubbed VMC-NQS. VMC calculations with a stochastic reconfiguration optimizer is equivalent to a rigorous Euclidean time projection method \cite{PhysRevB.61.2599, Stokes2020}, as long as the projection path can be spanned by the ansatz. The limited representation capability of conventional ansatz restricts the accuracy of the VMC method, which can be effectively addressed by NQS. A series of recent works showed that NQS can be effectively adapted to investigate various quantum many-body systems, such as atoms and molecules \cite{deepwf, Ferminet, vonglehn2023selfattentionansatzabinitioquantum, PauliNet, Symmetries, Laplacian, FermiSci}, condense matter \cite{chen_empowering_2024, luo_simulating_2024, teng_solving_2025, linteau_phase_2025, qian_describing_2025, li_emergent_2024, li_deep_2025, linteau_universal_2025}, atomic nuclei \cite{v2007,A6, Hidden,  Feynmannet,distill, YANG2022137587,cpl_42_5_051201}, ultra-cold Fermi gas \cite{Kim2024,PhysRevX.14.021030}, homogeneous electron gas \cite{elegas} and the crust of neutron stars \cite{PhysRevResearch.5.033062, crust}. Though VMC-NQS is conceptually straightforward, the integration of hidden nucleonic degrees of freedom \cite{Hidden} and neural-network backflow transformations \cite{Feynmannet} greatly enhances its accuracy in nuclear $ab$ $initio$ calculations, surpassing more sophisticated methods like hyperspherical-harmonics (HH) and AFDMC. This positions VMC-NQS as a promising method to push the boundaries of variational study of light hypernuclei beyond the limitations of SVM. 

In this work, we extend the scope of NQS to hypernuclei, filling the gap of investigating baryonic systems beyond protons and neutrons. Ground and excited states of $A\leq 5$ single-$\Lambda$ hypernuclei are calculated and benchmarked against SVM results. The NQS architecture is designed following Slater-Jastrow framework, with the bulk of baryonic correlations captured through backflow transformations\cite{PhysRevLett.122.226401} within the permutation-equivariant message-passing neural-network (MPNN) \cite{MPNN}. To reduce numerical errors for improved resolution of shallow $\Lambda$ potential and enable unified treatment of nucleons and hyperons, we developed the spinor grouping method to analytically integrate out isospin degrees of freedom.
Compared to molecules, the low degeneracy of energy levels in hypernuclei allows for the direct extraction of excited states by targeting the pure spin states, bypassing the high computational cost of the overlap penalty method \cite{Entwistle2023} and the natural excited state method \cite{FermiSci}. Guided by this requirement, a state-of-the-art spin purification scheme is proposed. Our VMC-NQS family outshines SVM in accuracy across all benchmarked nuclei. Through comparison of the $\Lambda$ separation energy of $^5_\Lambda$He with experimental results, we determine the optimal $\slashed{\pi}$EFT model. Utilizing this Hamiltonian, we investigate the energy spectrum of charge-symmetric $p$-shell hypernuclei up to $^{13}_\Lambda$C, exhibiting appreciable consistency with experimental measurements. 
A Gibbs-inspired sampling algorithm has been implemented to accelerate convergence to detailed balance. 
To achieve more robust and effective training, we have designed a novel learning rate scheduling strategy that enables finer-grained parameter updates near global minima. In this paper, we use $i$, $j$, $k$ to label nucleons and $n$, $m$, $l$ for baryon indices.

\section*{Results}

\subsection*{Energy benchmarks}
The properties of single-$\Lambda$ hypernuclei are governed by two distinct interactions, namely, the ones exclusively among the nucleons and those involving the extra hyperon. Similar to previous works\cite{Hidden, Feynmannet, A6, distill }, the former is modeled by leading-order (LO) $\slashed{\pi}$EFT, specifically the optimal model \enquote{o} as outlined in Ref.~\cite{pionless}, which is considered as \enquote{essential} elements of nuclear binding \cite{distill}. This model incorporates two-nucleon ($NN$) potential calibrated by reproducing the $np$ effective range expansions in $S/T = 0/1$ and $1/0$ channels, with an additional repulsive three-nucleon ($3N$) force to avoid Thomas collapse \cite{thomas} and reproduce the ground energies of heavier nuclei, adjusted to $^3$H binding energy. Additionally, the electromagnetic component is assumed to act solely between protons of finite size. 

For the latter one, we here leverage $s$-wave $\Lambda N$ and $\Lambda NN$ contact interactions described by LO $\slashed{\pi}$EFT
\begin{equation}
	\begin{aligned}
		V_{\Lambda N} &= \sum_{S} C_\lambda^S \sum_{i} \mathcal{P}_{S}(\Lambda i) e^{-\frac{\lambda^2}{4}r^2_{\Lambda i}},\\
		V_{\Lambda NN} &= \sum_{SI} D^{SI}_\lambda \sum_{i<j} \mathcal{Q}_{SI}(\Lambda ij) e^{-\frac{\lambda^2}{4}(r^2_{\Lambda i}+r^2_{\Lambda j})},
		\label{eq: Hamiltonian}
	\end{aligned}
\end{equation} with original low-energy constants (LECs) sourced from Ref. \cite{PhysRevC.106.L031001}.
\( \mathcal{P}_S \) and \( \mathcal{Q}_{SI} \) are projection operators act on \( \Lambda N \) pairs (spin \( S \)) and \( \Lambda NN \) triplets (spin \( S \), isospin \( I \)), respectively. The depth of $\Lambda N$ force is primarily fixed by fitting low-energy scattering length \cite{PhysRevC.106.L031001}. The LECs in $\Lambda NN$ force \cite{PhysRevC.106.L031001} are determined by reproducing the $\Lambda$ separation energies ($B_\Lambda$) of $^3_\Lambda$H$^{\frac{1}{2}^+}$, $^4_\Lambda$H$^{0^+}$ and $^4_\Lambda$H$^{1^+}$. In this work, we set the momentum cutoff $\lambda$ in the local Gaussian regulator in Eq. (\ref{eq: Hamiltonian}) to be 2 fm$^{-1}$. To facilitate comparison with existing SVM methods, we have also done calculations using the identical nuclear force as in Ref.~\cite{PhysRevC.106.L031001}, which is applicable only to $s$-shell atomic nuclei. In what follows, the combination of the former nuclear force and $YN$ interaction as in Eq. (\ref{eq: Hamiltonian}) is denoted as Hamiltonian \enquote{o}, with Hamiltonian \enquote{s} referring to that with the latter $s$-shell specific nuclear force. 
\begin{figure*}[htb]
	\centering
	\includegraphics[ width=0.7\textwidth]{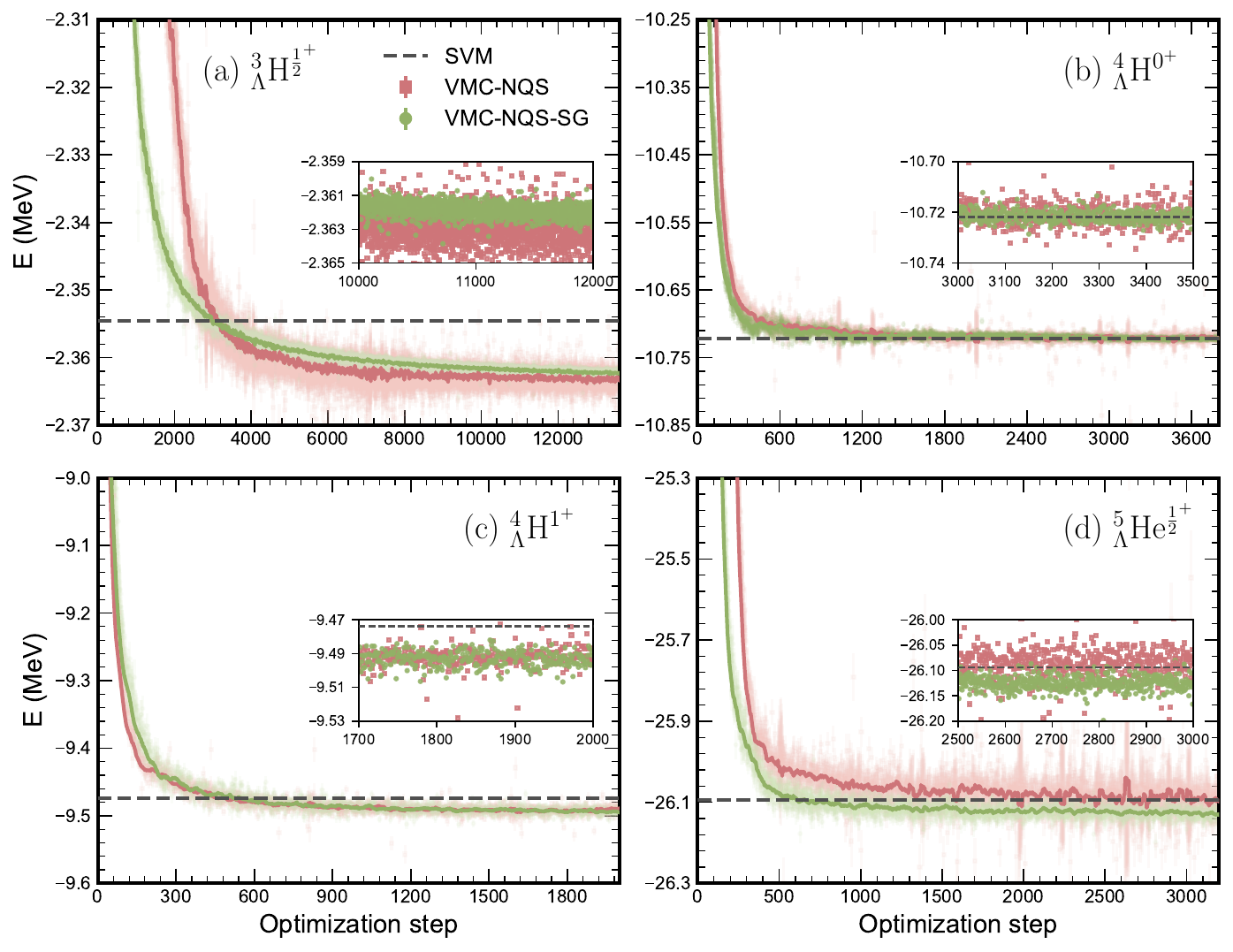}
	\caption{\parbox[t]{\textwidth}{FIG. 1. Energy convergence patterns of (a) $^3_\Lambda$H$^{\frac{1}{2}^+}$, (b) $^4_\Lambda$H$^{0^+}$, (c) $^4_\Lambda$H$^{1^+}$ and  (d) $^5_\Lambda$He$^{\frac{1}{2}^+}$, obtained with Hamiltonian \enquote{s}. Raw data points from VMC-NQS are shown as red solid squares, while green solid circles denote that with spinor grouping technique (VMC-NQS-SG).  The corresponding sliding curve and error bars are obtained by calculating the mean and standard deviation over a 25-step sliding window. Results of SVM method \cite{PhysRevC.106.L031001} are displayed by gray dashed lines. Inset zoomed regions highlight late-stage convergence behavior.
	}}
	\label{fig:energy convergence}
\end{figure*} 

Despite the remarkable success of NQS in accurately describing atomic nuclei, their extension to light hypernuclei presents unique challenges, as the additional $\Lambda$ hyperon resides in a characteristically shallow potential.
First of all, to model the exact hyperon halo, it is imperative that the statistical error is negligible relative to $B_\Lambda$ values. In QMC calculations, such a reduction in the statistical error demands a larger sample size, posing a challenge to resources. To address these issues, we present the spinor grouping (SG) method to analytically integrate out the conserved isospin degrees of freedom, as detailed in the Spinor grouping method subsection of the Methods. Additionally, the loss landscape of hypernuclei is notably intricate due to their near-degeneracy property, that is the energy differences between adjacent spin states are kept below $B_\Lambda$ values. A spin-unconstrained VMC-NQS training would easily get trapped in suboptimal local minima, corresponding to highly spin-contaminated states \cite{Symmetries}. Here, we directly enforce NQS for hypernuclei with $A\leq4$ to be eigenstates of $\bm{S}^2$ by means of spin projection operators. Further discussion on this spin symmetry enforcement can be found in Spin purification subsection of the Methods.

In this study, we investigate the spectrum of light single-$\Lambda$ hypernuclei using the baseline VMC-NQS method, as well as the SG enhanced version referred to as VMC-NQS-SG. Both of them are based on Slater-Jastrow ansatz, enhanced by an MPNN backflow transformation, dubbed SJ-BF. The expressive power of NQS is demonstrated by its comparison with existing stochastic variational method.
All assessments on the efficiency of SG method are carried out under strictly identical conditions, including the same number of samples ($N=40,000$), determinants ($K=3$) and the depth of MPNN ($L=3$). The Monte Carlo sampling is carried out through a Gibbs-inspired Metropolis-Hastings process (MH-Gibbs).
To achieve more robust and efficient training, all simulations in this paper adopt a novel learning rate schedule that steers the evolution of quantum states in Hilbert space along an optimally smooth trajectory. Full details of this schedule, together with the MH-Gibbs sampling and ansatz architecture can be found in the Optimization, Sampling and Neural-network quantum states subsections of the Methods. 

Fig. \ref{fig:energy convergence} shows the energy convergence for $^3_\Lambda$H$^{\frac{1}{2}^+}$,  $^4_\Lambda$H$^{0^+}$,  $^4_\Lambda$H$^{1^+}$ and   $^5_\Lambda$He$^{\frac{1}{2}^+}$ calculated using Hamiltonian \enquote{s}, plotted as a function of optimization steps. The existing SVM results from Ref. \cite{PhysRevC.106.L031001} are taken as the benchmark. For all examined hypernuclei, our VMC-NQS family, either the standard or the SG improved version, yields energies lower than or at least consistent with SVM results. The fact that our energies match or surpass those of SVM is particularly noteworthy, given that SVM benefits from having the analytical evaluation of correlated Gaussian basis matrix elements, making it a fairly accurate few-body
method.  This indicates that properly designed SJ-BFs are better approximators for baryonic systems than correlated
Gaussian bases. 

Moreover, it is palpable that across all panels in Fig. \ref{fig:energy convergence}, the variance obtained with VMC-NQS-SG method is substantially lower than that from original VMC-NQS calculation, albeit with a fixed sample size. This outcome is as expected, since the integral in isospin space is treated analytically rather than using probabilistic sampling. Of note is that, because SG method inherently aligns the dimensions of nucleons and hyperons, the NQS used in VMC-NQS calculations incorporates an additional dimensional alignment component and a larger Jastrow factor. Consequently, it has hundreds more trainable parameters compared to the SG variant, which should in principle augment its expressive capability and thereby provide a better energy upper limit.
For both $^3_\Lambda$H and $^4_\Lambda$H, the VMC-NQS and VMC-NQS-SG methods converge to highly consistent energies. For $^5_\Lambda$He, the SG energy is lower than the SVM and VMC-NQS results by more than 40 keV. These observations suggest that when the ansatz can achieve sufficient coverage of the vicinity of the exact state, the precision of expectation value estimation becomes the limiting factor for higher accuracy. This effect is accentuated in $^{5}_{\Lambda}\mathrm{He}$ due to its greatly expanded model space compared to $^{3}_{\Lambda}\mathrm{H}$ and $^{4}_{\Lambda}\mathrm{H}$, substantially amplifying sampling difficulties. Notably, for the calculation of the ground energy of $^5_\Lambda$He, we employ no spin purification scheme. Nevertheless, the VMC-NQS-SG method still yields an energy significantly lower than the SVM result. This is particularly remarkable given that a very recent preprint \cite{didonna2025hypernucleineuralnetworkquantum}, applying the standard VMC-NQS method to single-$\Lambda$ hypernuclei, failed to improve upon the SVM prediction for the ground energy of $^5_\Lambda$He. 
This finding once again highlight the effectiveness of SG method. 

\subsection*{$\Lambda$ separation energies, radii and spatial distributions of $s$-shell hypernuclei}
\begin{figure}[tbp]
	\includegraphics[width=0.49\textwidth]{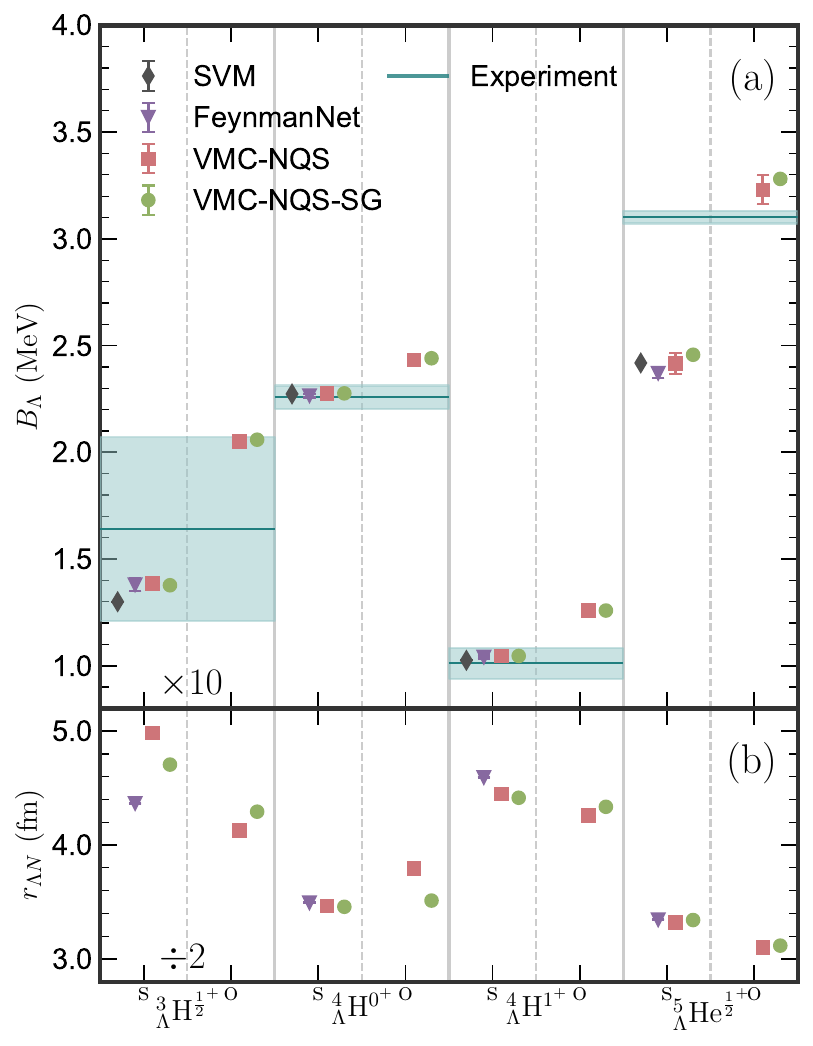}
	\caption{\parbox[t]{0.49\textwidth}{FIG. 2. Results of (a) $B_\Lambda$ and (b) $r_{\Lambda N}$ obtained from SVM \cite{PhysRevC.106.L031001}, FeynmanNet, VMC-NQS and VMC-NQS-SG calculations, presented as gray  diamonds, purple lower triangles, red squares and green  circles respectively. The results and error bars are taken as the mean and standard deviation over the last 100 iterations. The blue horizontal line and shaded area represent experimental measurements and uncertainties, taken from Ref.~\cite{eckert2021hypernuclides} with four-body results averaged.}}
	\label{fig:BL}
\end{figure}

The subfigure (a) in Fig. \ref{fig:BL} juxtaposes $B_\Lambda$ values resulting from SVM \cite{PhysRevC.106.L031001} and VMC-NQS variants for models \enquote{s} and \enquote{o}, with experimental measurements \cite{eckert2021hypernuclides} are displayed as reference lines. The results from FeynmanNet \cite{Feynmannet} based ansatz have also been provided.
In the absence of charge symmetry breaking (CSB) effects in our Hamiltonian, we use averaged $B_\Lambda$ experimental values for four-body hypernuclear systems. The specific figures and the convergence curves for Hamiltonian \enquote{o} are listed in the supplementary material. Similar to the case in model \enquote{s}, the SG technique significantly boosts both the accuracy and precision of energy estimation in VMC-NQS. With the SG method, we observe a reduction in variance by up to an order of magnitude, achieving one-thousandth level accuracy for the prediction of $B_\Lambda$ values. Given the VMC treatment for the full Hamiltonian across the complete space, it avoids the need for extrapolations in imaginary time, volume and model space, which are typically required in methods such as NLEFT and NCSM. All numerical errors in VMC calculations originate from statistical uncertainty and can be systematically managed through SG method and enlarging the sample size. This is particularly crucial for constraining the LECs of three-baryon interactions, which necessitate accurate determination of the $\Lambda$ separation energies.
\begin{figure}[tbp]
	\includegraphics[width=0.49\textwidth]{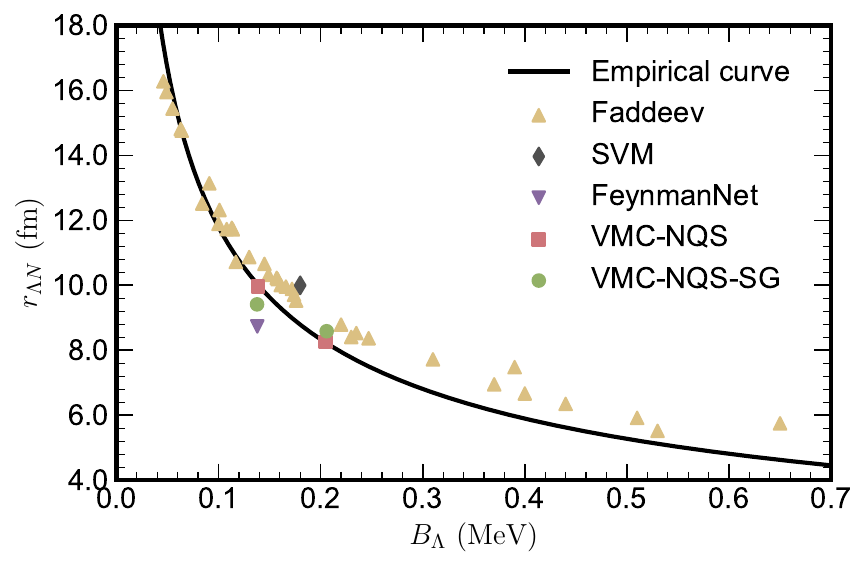}
	\caption{\parbox[t]{0.49\textwidth}{FIG. 3. The $r_{\Lambda N}$ dependence of $B_\Lambda$ of hypertriton. The black curve refers to the empirical relation. The yellow upper triangles, gray diamonds, red squares , green  circles and purple lower triangles are results from Faddeev \cite{Cobis_1997}, SVM \cite{Nemura1999StudyOL}, VMC-NQS, VMC-NQS-SG and FeynmanNet calculations, respectively. 
    %The error bars are not plotted because they are too small to be visible. 
    }}
	\label{fig:R}
\end{figure}

In the lower panel (b) of Fig. \ref{fig:BL}, we show the distance between $\Lambda$ and nuclear core, computed using VMC-NQS family, which is defined as \cite{Nemura1999StudyOL}
\begin{equation}
	\begin{aligned}
		r_{\Lambda N} = \sqrt{\frac{1}{A-1}\left\langle\Psi\left|\sum_i \left\|\bm{r}_{\Lambda}-\bm{r}_i\right\|^2\right|\Psi\right\rangle}.
	\end{aligned}
\end{equation}
The $\Lambda N$ distances hold particular significance in high-energy nuclear physics, because the coalescence model requires theoretical evaluations of hypernuclear radii for various $B_\Lambda$ values as input \cite{LIU2024138855,  PhysRevLett.134.022301}, which are challenging to measure experimentally.
For several nuclei, the predicted spatial extensions from the VMC-NQS and VMC-NQS-SG methods show minor deviations. This is reasonable since the convergence rate of radii is typically half that of energies. In light single-$\Lambda$ hypernuclei, the halo structure further constrains the energy difference between states with widely disparate radii to remain below $B_\Lambda$, imposing additional challenges for determining the exact $r_{\Lambda N}$ in the ground states. Despite these difficulties, our VMC-NQS variants still yield fairly accurate $r_{\Lambda N}$ results.

The hypertriton, featuring a loosely bound $\Lambda$ hyperon around a deuteron ($d$) core, is the lightest known nuclear halo system. By simplifying the $\Lambda d$ force with an attractive quadratic potential, one can write down a phenomenological relation between $r_{\Lambda N}$ and $B_\Lambda$ independent of the specific Hamiltonian \cite{LIU2024138855},
\begin{equation}
	\begin{aligned}
r_{\Lambda N} \approx \frac{\hbar c}{\sqrt{4 \mu B_\Lambda}},
	\end{aligned}
\end{equation}
where $\mu=M_\Lambda M_d/(M_\Lambda + M_d)$ is the reduced mass. As shown in Fig. \ref{fig:R}, the $\Lambda d$ distances obtained from VMC-NQS family are in excellent agreement with this empirical curve.
Our results are further supported by Faddeev \cite{Cobis_1997}, SVM \cite{Nemura1999StudyOL} and FeynmanNet calculations, showing consistent alignment. This confirms that the VMC-NQS family can accurately describe hypernuclear properties beyond just eigenenergies.
%The binding energy of hypertriton modeled by the \enquote{s} Hamiltonian is 2.3632(11) MeV and 2.3623(3) MeV, obtained with VMC-NQS and the SG improved version separately, markedly higher than the SVM result, 2.3546 MeV. 

\begin{figure}[tbp]
	\includegraphics[width=0.49\textwidth]{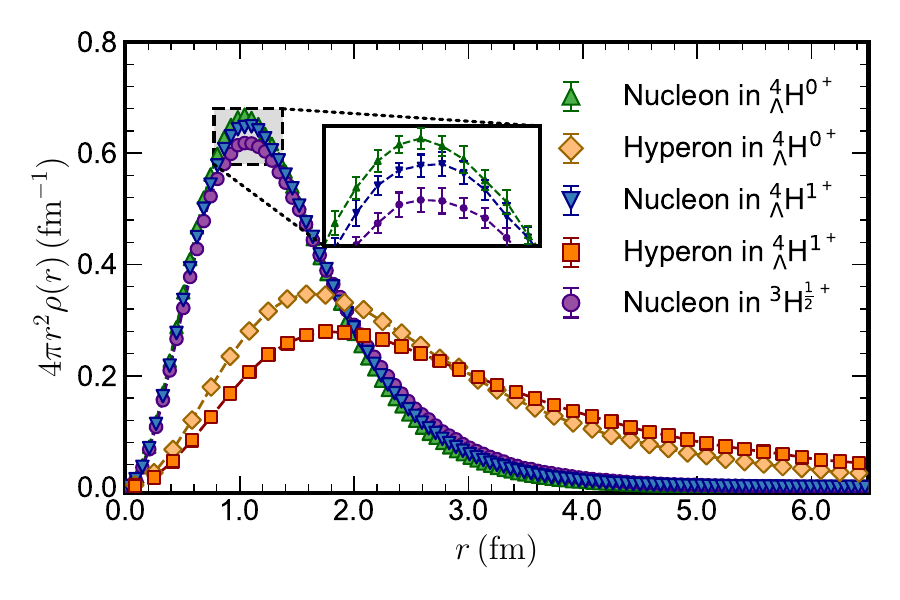}
	\caption{\parbox[t]{0.49\textwidth}{FIG. 4. The spatial distributions of $^3$H$^{\frac{1}{2}^+}$ , $^4_\Lambda$H$^{0^+}$ and $^4_\Lambda$H$^{1^+}$ obtained with VMC-NQS-SG method. Green upper triangles, blue lower triangles and purple circles denote the point-nucleon distributions. The $\Lambda$ orbits are presented as yellow  diamonds and orange  squares for ground and excited states.}}
	\label{fig:distribution}
\end{figure}

To further demonstrate that NQS provides access to all nuclear eigenstate properties, we also investigate the spatial distribution of nucleons and $\Lambda$ hyperon
\begin{equation}
	\begin{aligned}
		&\rho_N(r) = \frac{1}{4\pi r^2} \frac{\langle \Psi | \sum_{i} \delta(\left\|\bm{r}_i-\bm{R}_{\text{n.c.}}\right\| - r) | \Psi \rangle}{(A-1)\langle \Psi | \Psi \rangle},\\
		&\rho_\Lambda(r) = \frac{1}{4\pi r^2} \frac{\langle \Psi |\delta(\left\|\bm{r}_\Lambda-\bm{R}_{\text{n.c.}}\right\| - r) | \Psi \rangle}{\langle \Psi | \Psi \rangle},
	\end{aligned}
\end{equation}
where $\bm{R}_{\text{n.c.}}$ is the nuclear core coordinate.
Fig. \ref{fig:distribution} depicts $\rho_N(r)$ and $\rho_\Lambda(r)$ of $^3$H$^{\frac{1}{2}^+}$ , $^4_\Lambda$H$^{0^+}$ and $^4_\Lambda$H$^{1^+}$ obtained from the VMC-NQS-SG method.
One can observe the shrinkage effect, where the nuclear core experiences slight compression as the $\Lambda$ hyperon approaches, which has been extensively discussed over the past decades \cite{Nemura1999StudyOL, PhysRevC.59.2351}. Excitations predominantly result from variations in the hyperon orbits. This behavior is in line with the Gamow shell model calculation assumption of a frozen nuclear core \cite{LI2025139708}, as the $YN$ interaction is much weaker than that among nucleons. These outcomes once again corroborate the correctness of NQS in representing the hypernuclear wavefunctions.
\subsection*{Extrapolation to $p$-shell hypernuclei}
\begin{table}[tbp] 
\caption{\parbox[t]{0.49\textwidth}{Table 1. Numerical results for $\Lambda$ separation energy and $\bm{S}^2$ expectations of $p$-shell hypernuclei from VMC-NQS-SG method. The results and error bars are taken as the mean and standard deviation over the last 100 iterations.  The experimental results are taken from Ref.~\cite{eckert2021hypernuclides}.}}
\begin{tabular}{p{0.7cm} c c c |c}
\toprule
 & $B_\Lambda$ (MeV) & \text{Exp.} (MeV)& $B_\Lambda$ / \text{Exp.} &  \multicolumn{1}{c}{$\langle\bm{S}^2\rangle$}\\ 
\midrule
$^7_\Lambda$Li$^{\frac{1}{2}^+}$ & 5.374(60) & 5.619(60) & 0.96 & 0.756(1) \\
$^7_\Lambda$Li$^{\frac{3}{2}^+}$ & 4.640(60) & 4.927(60) & 0.94 & 3.7529(8)   \\
$^9_\Lambda$Be$^{\frac{1}{2}^+}$ & 5.485(54) & 6.614(72) & 0.83 & 0.772(2)  \\
$^{11}_\Lambda$B$^{\frac{1}{2}^+}$ & 8.127(87) & 8.767(58) & 0.93 & 0.789(2)  \\
$^{11}_\Lambda$B$^{\frac{3}{2}^+}$ & 7.952(71) & 8.263(58) & 0.96 & 3.791(2)  \\
$^{13}_\Lambda$C$^{\frac{1}{2}^+}$ & 10.64(11) & 11.80(16) & 0.90 & 0.787(3)  \\
\bottomrule
\label{p}
\end{tabular}
\end{table}
In this subsection, we aim to investigate the prediction ability of LO $\slashed{\pi}$EFT into heavier single-$\Lambda$ hypernuclei. Given that the LECs in the $\Lambda NN$ force are adjusted to fit experimental $B_{\Lambda}$ values only for $A \leq 4$ hypernuclei, the predictive power lies in its extrapolation for heavier hypernuclei ($A > 4$), such as ${}^{5}_{\Lambda}\mathrm{He}$. According to the (a) panel of Fig. \ref{fig:BL}, the model \enquote{o} yields a separation energy for $^5_\Lambda$He that is much closer to experimental data than model \enquote{s}. This outcome is as expected because the nuclear force in model \enquote{s} features a fixed momentum cutoff for $S/T=0/1, 1/0$ channels. As pointed out in Ref. \cite{SVMPRL}, to obtain the correct $B_\Lambda$ value for $^5_\Lambda$He, the cutoff $\lambda$ must be extended to infinity. However, within $\slashed{\pi}$EFT, $p$-shell nuclei would break up into sub-clusters once the cutoff scale exceed a critical value \cite{multifermion}. This problem is addressed in model \enquote{o} by determining the LECs in different $s$-wave channels independently.
%As a result, is limited in $s$-shell atomic nuclei, which results in a binding energy of approximately 23 MeV for $^4$He. 
As a result, unlike the $s$-shell specific nuclear force in model \enquote{s}, that in model \enquote{o} can provide fairly great descriptions for atomic nuclei with mass number up to 40 \cite{pionless}, and is considered as \enquote{essential} elements for nuclear binding \cite{distill}. Therefore, we adopt Hamiltonian \enquote{o} to model $p$-shell hypernuclei. Because of the superior performance of the SG method shown in the Energy benchmark subsection, that it can effectively boost both accuracy and precision while saving computational resources, we report exclusively results from VMC-NQS-SG calculations, with $K=6$. Similar to the case of $^5_\Lambda$He, the exponentially growing model space of $p$-shell hypernuclei greatly amplifies the effectiveness of SG method, making the SG-improved predictions significantly surpass the VMC-NQS energies. Further comparisons can be found in the supplementary material. 

In Table \ref{p}, we summarize the predicted $B_\Lambda$ and the resulting $\bm{S}^2$ expectations for low-lying states of charge-symmetric $p$-shell single-strangeness hypernuclei with mass number up to 13. For all examined nuclei, except for $^{9}_\Lambda$Be, model \enquote{o} shows fairly accurate predictions of the $B_\Lambda$ values, achieving a 90\% agreement with experimental data \cite{eckert2021hypernuclides}. The $B_\Lambda$ uncertainty propagates from the quadrature sum of energy uncertainties in the hypernucleus and its nuclear core.  The binding energies of $^{10}$B and $^{12}$C are 61.78(1) MeV and 87.75(7) MeV respectively, provided by the FeynmanNet group \cite{Feynmannet}. Since $^8$Be is a resonance system composed of two $\alpha$ clusters, neither our team nor the FeynmanNet group has been able to identify a stable bound state. Hence, we here use twice the ground energy of $^4$He to substitute for the nuclear core energy of $^9_\Lambda$Be. This approximation neglects the Coulomb repulsive effect between two $\alpha$ clusters, which is hard to identify through VMC based calculation. This accounts for the outlier status of $^{9}_{\Lambda}\mathrm{Be}$ in the model predictions compared to experimental results. Nevertheless, similar to the extra neutron in $^9$Be \cite{PhysRevLett.134.162503}, the $\Lambda$ hyperon that moves around in a \enquote{molecular orbital} binds the two $\alpha$ cluster together. 

Given the absence of tensor force in our nuclear force, the ordering of $1^+$ and $3^+$ states of $^{10}$B is inverted \cite{PhysRevLett.89.182501}. As a consequence, the ground state of $^{10}$B obtained in our model is $1^+$. Therefore, we shift our focus from the experimental ground state to the hypernuclear states $^{11}_\Lambda$B$^{\frac{1}{2}^+}$ and $^{11}_\Lambda$B$^{\frac{3}{2}^+}$, both of which share the $1^+$ nuclear core configuration obtained in our model, coupled with hyperon in $s$- and $p$-wave orbits respectively. 

The resulting $\langle\bm{S}^2\rangle$ can be found in the rightmost column of Table \ref{p}. For the heavier systems, we use the penalty method as in Ref~\cite{FermiSci} to eliminate spin contamination, which is detailed in the Spin purification subsection of Method. The $\omega_s$ is set to be 3 MeV for the calculations of $^7_\Lambda$Li$^{\frac{1}{2}^+}$ and $^{11}_\Lambda$B$^{\frac{1}{2}^+}$ to widen the significantly small energy gap between ground and first excited states, and 1 MeV for the rest nuclei. 
%Although we have trained the NQS for sufficient long time, we still found a certain degree of spin contamination in the final obtained states.
The impact of this spin contamination can be assessed by performing diagonalization of the resulting state. Taking $^{11}_\Lambda$B$^{\frac{1}{2}^+}$ as an example, similar to the multi-channel method in NLEFT, we assume the $\frac{1}{2}^+$ and $\frac{3}{2}^+$ are dominant components in the resulting state. Thus, the proportion of the contribution from exact ground state, $a$, can be evaluated by solving such linear equation
\begin{equation}
	\begin{aligned}
	0.75 a + 3.75 (1-a) = \langle\bm{S}^2\rangle,
		\label{eq:proportion}
	\end{aligned}
\end{equation}
which results in $a=0.987$. Then we can reproduce the true ground energy $E_{\text{g.s.}}$ 
\begin{equation}
	\begin{aligned}
	E_{\text{g.s.}} = \frac{E_{1/2}-(1-a)E_{3/2}}{a},
		\label{eq:proportion}
	\end{aligned}
\end{equation}
in which the $E_{1/2}$ and $E_{3/2}$ denote the obtained ground and excited state energy from VMC-NQS-SG calculations. The corrected $B_\Lambda$ value for $^{11}_\Lambda$B$^{\frac{1}{2}^+}$ is 8.129 MeV, with marginal refinement compared to the previous result. This indicates that including spin symmetry enforcement scheme can effectively purify the obtained eigenstates, given that with standard energy minimization, we find states with $a\approx0.6$ for both $^7_\Lambda$Li$^{\frac{1}{2}^+}$ and $^{11}_\Lambda$B$^{\frac{1}{2}^+}$. 

As seen in Table \ref{p}, excluding $^9_\Lambda$Be, the agreement between model \enquote{o} predictions and experimental values exhibits a slight downward trend with increasing mass number, ranging from 96\% for $^7_\Lambda$Li$^{\frac{1}{2}^+}$ down to 90\% for $^{13}_\Lambda$C$^{\frac{1}{2}^+}$. This is reasonable because of two limitations of this Hamiltonian. Firstly, the LECs for three-baryonic force in \enquote{o} are fitted to experimental values for nuclei with mass number up to 4. Therefore, the properties of heavier systems can be predicted solely by extrapolation, which could lead to severe deviations from the physical world. One way out of this dilemma is to fit the LECs while taking into account the heavy nuclei and carry out theoretical predictions through interpolation. A work very recently appears in the preprint website \cite{didonna2025hypernucleineuralnetworkquantum} constrains the LECs with the experimental results of $^{16}_\Lambda$O incorporated, thereby enhancing the accuracy of LO$\slashed{\pi}$EFT in describing $p$-shell hypernuclei. On the other hand, the lack of tensor force and spin-orbital effect inherently limits the prediction power of LO$\slashed{\pi}$EFT, which fails to reproduce the experimental spin value of the ground state of $^{10}$B \cite{PhysRevLett.89.182501} and $^{11}_\Lambda$B. The importance of these higher order terms is validated by a recent NLEFT work \cite{niu2025signproblemfreenuclearquantummonte}, in which the correct shell structure and binding energies from $^4$He to $^{132}$Sn are excellently reproduced with a simple SU(4) model plus spin-orbital coupling.

\section*{Discussion}
This work is motivated by the immense potential of neural-networks in representing variational quantum states. We propose the SG method to analytically integrate out the intrinsic isospin degrees of freedom, which enables the uniform treatment of nucleons and hyperons. By rigorously calculating a subset of the integrals, the SG method effectively reduces the statistical errors in Monte Carlo simulations, achieving a precision of less than 4\textperthousand for \( B_\Lambda \) calculations of light hypernuclei, and even less than 1\textperthousand for several nuclei. We further introduce the spin symmetry enforcement into nuclear VMC-NQS approach, which markedly accelerates the convergence speed. Together with a carefully designed MPNN based SJ-BF hypernuclear ansatz, the integration of SG method and spin purification techniques into VMC-NQS family enables its results to not only match but also surpass the energies obtained by SVM, a fairly accurate few-body method routinely employed in hypernuclei calculations. A Gibbs-inspired sampling algorithm is proposed to improve sampling efficiency. We have concurrently augmented the SR minimizer by confining quantum state variations at each optimization step to remain within a specified threshold, remedying the intermittent numerical instability.

By the highly accurate few-body calculations of light single-$\Lambda$ hypernuclei, we identify the optimal $s$-shell based $\slashed{\pi}$EFT model and extrapolate it to $p$-shell hypernuclei. We calculate the $B_\Lambda$ values of charge-symmetric hypernuclei with $A\leq 13$, exhibiting a 90\% agreement with experimental measurements. The effectiveness of SG method is further highlighted in our calculations for $p$-shell hypernuclei, in which the SG-improved VMC-NQS energies are significantly lower than the original version. The importance of including spin symmetry enforcement is further emphasized by a significant reduction in the spin contamination ratio in $^7_\Lambda$Li$^{\frac{1}{2}^+}$ and $^{11}_\Lambda$B$^{\frac{1}{2}^+}$. It is noteworthy that, as shown in Table \ref{p}
, that spin contamination becomes increasingly severe with higher mass numbers. This arises because the heavier nuclei possess more high spin states, which would remain in the final states through simple energy optimization. 

In the future, we aim to improve the performance of our current MPNN based neural-network ansatz by explicitly encoding mean-field information. As reported by FeynmanNet \cite{Feynmannet} and PauliNet \cite{PauliNet} groups, building in major shell structure and Hartree-Fock solutions can provide highly accurate results for larger nuclei and molecule systems, while being computationally efficient. This is reasonable because the dimension of the solution space grows exponentially with the number of particles. Therefore, it is much more challenging for a variational ansatz to represent the full space for heavier nuclei, even when using neural-networks. A smarter way out is to roughly locate the physical state using mean-field guidance, and explore the neighborhood via neural-network based backflow transformation. We anticipate this will sidestep the need for a great amount of variational parameters when providing lower energy caps for heavier atomic nuclei and hypernuclei.

Given our VMC-NQS-SG method currently yields the most accurate results, at least for light single-$\Lambda$ hypernuclei, it is well suited for determining the LECs in $\Lambda NN$ interaction and exploring CSB effects. Looking ahead, we are expecting to include multiple $\Lambda$ and $\Sigma$ states, a straightforward extension within the SG framework. This approach paves the way for more in-depth $ab$ $initio$ studies of hypernuclei with heavier quarks, as the SG paradigm offers a unified manner for incorporating degrees of freedom into VMC-NQS calculations.
\section*{Method}
\subsection*{Spinor grouping method}

%NQS are trainable mappings from many-body eigenstates to complex amplitudes, with essential symmetries imposed to reduce the redundant Hilbert space. Following the traditional NQS framework, the $\Lambda$ states can be incorporated by extending the input of the ansatz. For single-$\Lambda$ hypernuclei with mass number $A$, the NQS is a function $\Psi$: ($\bm{r}_i,s_i,t_i$)$^{A-1}\times(\bm{r}_\Lambda,s_\Lambda)\mapsto\mathbb{C}$, where $\bm{r}_{i(\Lambda)}\in\mathbb{R}^3$ and $s_{i(\Lambda)}\in\{1, -1\}$ denotes single-baryon spatial and spin degrees of freedom respectively. $t_i\in\{1, -1\}$ indicates the z-projection of isospins of the $A-1$ nucleons
%comprising the nuclear core. Whereas, since the $\Lambda$ state does not contain isospin degrees of freedom, operators in isospin space cannot directly interfere with $\Lambda$. Isospin operators can only affect $\Lambda$ indirectly by first altering the position-spin states of nucleons, which then influence position-spin states of $\Lambda$ through $YN$ interaction. However, if we simply create a blackbox mapping that mixes nucleons and hyperons, it would introduce a direct pathway from isospins to $\Lambda$ states. In a nutshell, the misalignment of particle degrees of freedom can introduce non-physical components into the NQS. The calculation of hypernuclei therefore requires a more delicate treatment. In this paper, we present spinor grouping (SG) method to resolve this issue. 

In SG formalism, the isospin degrees of freedom are analytically integrated out. This is achieved by constructing the trial state for nuclear core with isospins explicitly labeled. For atomic nucleus with mass number $A$ and proton number $Z$, we consider the trial state
\begin{equation}
	\begin{aligned}
		\left|\Psi\right\rangle = \sum_{\{\mathfrak{P}\}} \eta(\mathfrak{P}) \mathfrak{P} \left|\psi_1\right\rangle_1 \left|\psi_2\right\rangle_2 \ldots \left|\psi_A\right\rangle_A \left|\mathbf{p}\right\rangle_1 \ldots \left|\mathbf{n}\right\rangle_A,
		\label{SG-NQS}
	\end{aligned}
\end{equation}
where $\mathfrak{P}$ stands for permutations acting on particle indices, $\eta(\mathfrak{P})$ the parity of $\mathfrak{P}$. We use $\left|\mathbf{p}\right\rangle$ and $\left|\mathbf{n}\right\rangle$ to denote protons and neutrons respectively. Such restriction on Hilber space is reasonable because isospin is conserved when weak interactions are not involved. The residual states $\left|\psi_i\right\rangle$ are 
entangled states of nucleons in position-spin space, required to be permutation equivariant with nucleon indices. We now define following notations
\begin{equation}
	\begin{cases}
		|\Phi\rangle = |\psi_1\rangle_1 |\psi_2\rangle_Z \ldots |\psi_A\rangle_A, \\
		|\mathfrak{P}\rangle = \mathfrak{P} |\mathbf{p}\rangle_1 \ldots |\mathbf{p}\rangle_Z |\mathbf{n}\rangle_{Z+1} \ldots |\mathbf{n}\rangle_A, \\
		|\mathfrak{I}\rangle = |\mathbf{p}\rangle_1 \ldots |\mathbf{p}\rangle_Z |\mathbf{n}\rangle_{Z+1} \ldots |\mathbf{n}\rangle_A,
	\end{cases}
\end{equation}
and integrate out the isospins. Let us consider the projection of $\left|\Psi\right\rangle$ onto an arbitrary isospin state $|\mathfrak{P}_i\rangle$
\begin{equation}
	\begin{aligned}
		\langle \mathfrak{P}_i | \Psi \rangle &= \sum_{\{\mathfrak{P}\}} \eta(\mathfrak{P}) \langle \mathfrak{P}_i | \, \mathfrak{P} (|\Phi\rangle \otimes |\mathfrak{I}\rangle) \\
		&= \sum_{\{\mathfrak{P}\}} \eta(\mathfrak{P}) \langle \mathfrak{I} | \, (\mathfrak{P} |\Phi\rangle \otimes |\mathfrak{P}_i^{-1} \mathfrak{P}\rangle) \\
		%&= \eta(\mathfrak{P}_i) \mathfrak{P}_i \left( \sum_{\{\tilde{\mathfrak{P}}\}} \eta(\tilde{\mathfrak{P}}) \langle \mathfrak{I} | \, \tilde{\mathfrak{P}} (|\Phi\rangle \otimes |\mathfrak{I}\rangle) \right) \\
		&= \eta(\mathfrak{P}_i) \mathfrak{P}_i \langle \mathfrak{I} | \Psi \rangle.
		\label{project}
	\end{aligned}
\end{equation}
That is, we equate the permutations of isospins to that of the residual position-spin states.

We shall then describe how expectations of observables are computed with SG method. If the observable of interest, denoted as $O$, commutes with isospin, we are allowed to formally re-express it as a summation of effective permutations $\mathfrak{P}_\alpha$ and residual operators $R_\alpha$
\begin{equation}
	\begin{aligned}
		O = \sum_{\{\mathfrak{P}_\alpha\}} R_\alpha \mathfrak{P}_\alpha.
		\label{effective permutation}
	\end{aligned}
\end{equation}
Here $R_\alpha$ acts only on position-spin space. This gives the expectation of $O$ as
\begin{equation}
	\begin{aligned}
		\langle \Psi | O | \Psi \rangle &= \sum_{\{\mathfrak{P}\}} \sum_{\{\mathfrak{P}_\alpha\}} \langle \Psi | R_\alpha \mathfrak{P}_\alpha | \mathfrak{P} \rangle \langle \mathfrak{P} | \Psi \rangle \\
		&= \sum_{\{\mathfrak{P}_\alpha\}} \sum_{\{\mathfrak{P}\}} \eta^2(\mathfrak{P}) \mathfrak{P} \langle \Psi | R_\alpha | \mathfrak{P}_\alpha \rangle \mathfrak{P} \langle \mathfrak{I} | \Psi \rangle \\
		&= C_A^Z \left( \sum_{\mathfrak{P}_\alpha} \langle \Psi | R_\alpha | \mathfrak{P}_\alpha \rangle \right) \langle \mathfrak{I} | \Psi \rangle\\
		&= C_A^Z \left( \sum_{\mathfrak{P}_\alpha} \eta(\mathfrak{P}_\alpha)\mathfrak{P}_\alpha\langle \Psi | R_\alpha | \mathfrak{I}\rangle \right) \langle \mathfrak{I} | \Psi \rangle.
		\label{observable}
	\end{aligned}
\end{equation}
One can obtain the normalization factor for $| \Psi \rangle$ by substituting $O$ with identity, and this gives
\begin{equation}
	\begin{aligned}
		\langle \Psi | \Psi \rangle 
		= C_A^Z |\langle \mathfrak{I} | \Psi \rangle|^2.
		\label{norm}
	\end{aligned}
\end{equation}
The expectation of any observable, such as Hamiltonian, can be determined by combining Eq. (\ref{observable}) and (\ref{norm}).

%While evaluating the contribution of $NN$ interaction to the local energy, our SG method only requires summing over effective permutations rather than iterating through all nucleon pairs. 
\begin{figure}[tbp]
	\includegraphics[width=0.49\textwidth]{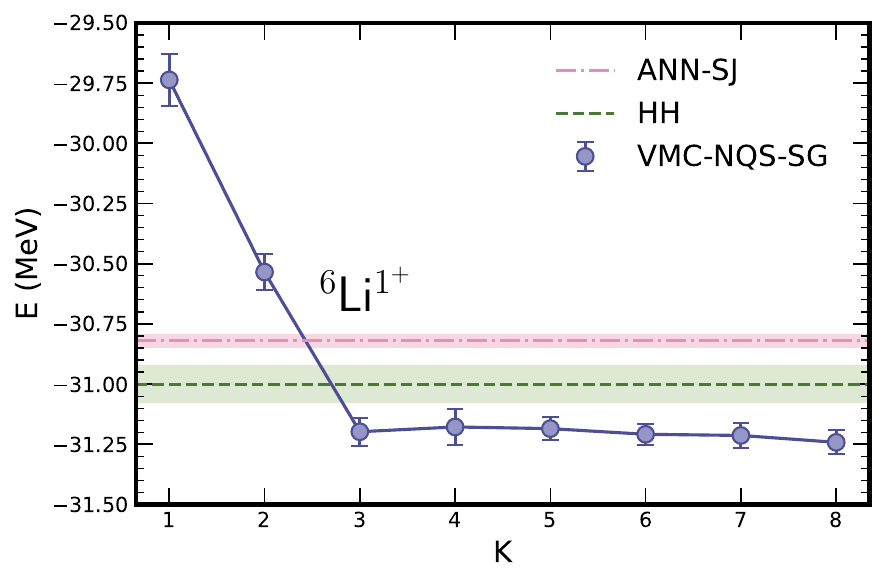}
	\caption{\parbox[t]{0.49\textwidth}{FIG. 5. Ground state energy of $^6$Li computed with VMC-NQS-SG, as a function of determinant number $K$, denoted by solid blue circles with error bars. The results and error bars are taken as the mean and standard deviation over the last 100 iterations. Results from the artificial neural network with Slater-Jastrow (ANN-SJ) ansatz and
			HH method \cite{A6} are shown as pink dash-dotted lines and green dashed lines, with shaded area as the error bars. }}
	\label{fig:6Li}
\end{figure}
With the SG technique, protons and neutrons can be effectively treated as non-identical particles, and we only compute the block-diagonalized determinants, such as for sampling and local energy calculations. These would considerably reduce computational complexity as the mass number increases. Furthermore, by analytically integrating out isospins instead of using Monte Carlo sampling, the SG method can effectively reduce memory usage and enhance numerical accuracy. %In practical calculation, we use multiple configurations to represent the nuclear quantum states, with each configuration taking the form of Eq. (\ref{SG-NQS}). 

It should be noted that while the a priori assignment of protons and neutrons in Eq. (\ref{SG-NQS}) suppresses potential isospin correlation, this limitation is systematically remediable through multi-configuration expansions and does not truncate the physical Hilbert space. To prove this point, we consider an arbitrary nucleonic state $\sum_{\{\mathfrak{P}\}}\eta(\mathfrak{P})\mathfrak{P}|\tilde{\Phi}\rangle$, as the antisymmetrization of an $A$ particle state $|\tilde{\Phi}\rangle$ in position-spin-isospin space.
We then expand the isospin sector in a complete basis and integrate out the isospin degrees of freedom
\begin{align}
	\sum_{\{\mathfrak{P}\}}\eta(\mathfrak{P})\mathfrak{P}|\tilde{\Phi}\rangle&=	\sum_{\{\mathfrak{P}\}}\eta(\mathfrak{P})\mathfrak{P}\left(\sum_{\{\mathfrak{P}_k\}}\langle\mathfrak{P}_k|\tilde{\Phi}\rangle\otimes|\mathfrak{P}_k\rangle\right)\nonumber\\
	%	&=\sum_{\{\mathfrak{P}_k\}}\sum_{\{\mathfrak{P}\}}\eta(\mathfrak{P})\mathfrak{P}\mathfrak{P}_k\left(\eta(\mathfrak{P}_k)\mathfrak{P}^{-1}_k\langle\mathfrak{P}_k|\tilde{\Phi}\rangle\right)\otimes|\mathfrak{I}\rangle\\
	&=\sum_{\{\mathfrak{P}_k\}}\sum_{\{\mathfrak{P}\}}\eta(\mathfrak{P})\mathfrak{P}\left(\eta(\mathfrak{P}_k)\mathfrak{P}^{-1}_k\langle\mathfrak{P}_k|\tilde{\Phi}\rangle\right)\otimes|\mathfrak{I}\rangle\nonumber\\
	&=\sum_{k}\sum_{\{\mathfrak{P}\}}\eta(\mathfrak{P})\mathfrak{P}|\Phi^{(k)}\rangle\otimes|\mathfrak{I}\rangle.
	\label{Expassion}
\end{align}
This indicates that the full set of $\frac{A!}{Z!(A-Z)!}$ configurations can span the entire Hilbert space. Yet far fewer suffice, as the residual states $|\Phi^{(k)}\rangle$ often coincide.
For instance, nuclei well described by the shell model require only one configuration, as demonstrated via Eq. (\ref{project}). Even for open-shell nuclei, few configurations are required as nucleons are already sufficiently entangled in position-spin space. To gauge the minimal isospin basis size required empirically, we monitor the ground energy of $p$-shell nucleus $^6$Li as a function of the number of configurations, denoted as $K$, in Fig. \ref{fig:6Li}. Results from the artificial neural network with Slater-Jastrow (ANN-SJ) ansatz and
HH method \cite{A6} are presented for comparison. One can see that just two determinants can yield a bound state for $^6$Li. This is a highly non-trivial result, given conventional VMC calculations, employing two- and three-body Jastrow correlations typically fail to bind $^6$Li below the 
$\alpha+d$ threshold, as pointed out in Ref.~\cite{A6} Upon increasing 
$K$ to three, the ground energy reaches a convergence to values lower than ANN-SJ and HH energy by approximately 450 keV and 250 keV respectively. This suggests three configurations are sufficient enough to describe the complex $\alpha+d$ cluster structure of $^6$Li. We therefore employ three determinants for light hypernuclei with $s$-shell nuclear cores. Note that, as demonstrated by experiments, our MPNN based SJ-BF ansatz requires minimal extra memory use when increasing determinants, enabling calculations for heavier hypernuclei with larger basis sets.

\subsection*{Spin purification}
Since the LO $\slashed{\pi}$EFT models do not contain tensor and spin-orbit forces, they share eigenstates with total spin square $\bm{S}^2$ and total orbital angular momentum square $\bm{L}^2$.
%In this work, the systems we focus on all have orbital angular momentum well approximated by $l=0$ \cite{Nemura1999StudyOL}. 
Therefore, different states of single-$\Lambda$ hypernuclei can be labeled by spin value $s$ and parity $\pi$. Parity can be naturally preserved by rewriting the trial wavefunction as $\Psi^P = (1 + \pi P)\Psi$ \cite{Hidden,Feynmannet,Kim2024}, with $P$ being the space inversion operator. However, the enforcement of NQS to be an eigenstate of $\bm{S}^2$ with eigenvalue $s(s+1)$ remains elusive. While fixing the $z$-components of total spin, $s_z$, is convenient during sampling, spin contamination from higher spin states remains inevitable. This manifests strongly in light hypernuclear calculations due to the weak spin-dependent $YN$ interaction, which impedes the self-emergence of stable singlets during training. With standard energy minimization, it takes about $10^5$ and $10^4$ steps to lower the energy below -2.3 MeV and -10.7 MeV for the ground state of $^3_\Lambda$H and $^4_\Lambda$H respectively, with the spin expectation around 1.6 and 0.01. This unsatisfying convergence behavior can be understood by noting that VMC with SR optimizer constitutes a sign-problem-free analogue of the Euclidean time projection method, where the convergence rate is primarily governed by the energy gap between the ground state and the first excited state. In the case of single-$\Lambda$ hypernuclei, the $B_\Lambda$ values serve as a natural cap on excitation energies, thus restricting the energy convergence pace.

\begin{figure}[tbp]
	\includegraphics[width=0.49\textwidth]{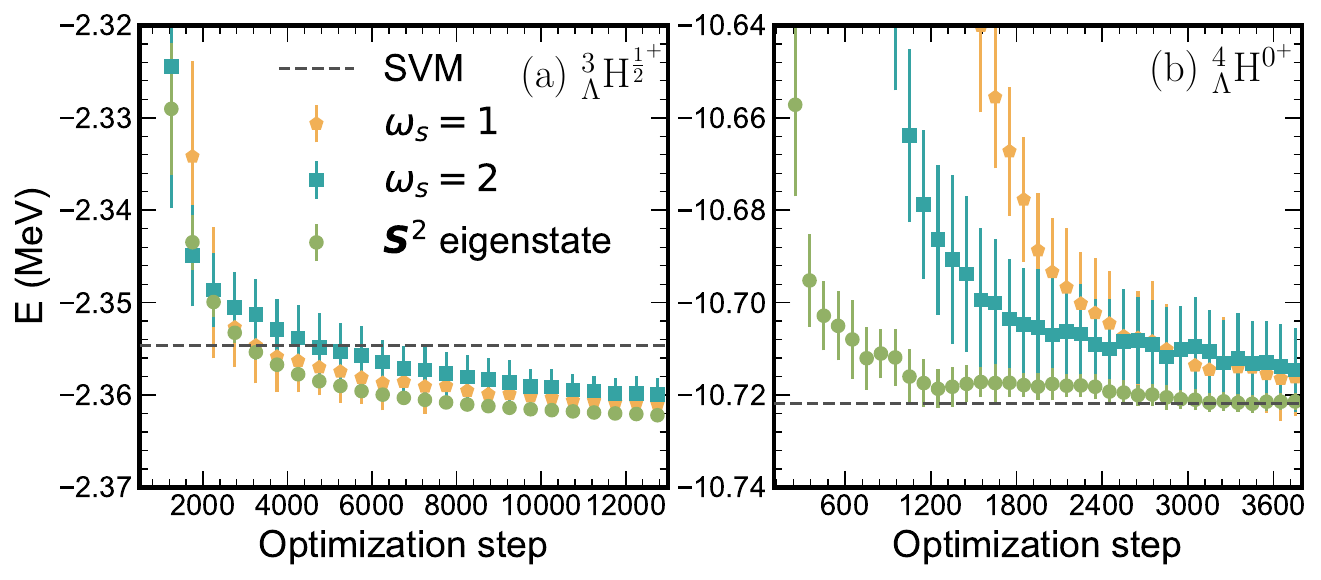}
	\caption{\parbox[t]{0.49\textwidth}{FIG. 6. comparison of energy convergence, using different spin purification schemes. For clarity, we bin the raw data points every 500 steps for (a) $^3_\Lambda$H$^{\frac{1}{2}^+}$ and 100 steps for (b) $^4_\Lambda$H$^{0^+}$ respectively. Then, we present the mean and standard deviation of the data within each bin. In both panels, green solid circles show convergence with spin enforced via the projection operator, while orange pentagons ($\omega_s$ = 1 MeV) and blue squares ($\omega_s$ = 2 MeV) show those from the penalty method. Results of SVM method \cite{PhysRevC.106.L031001} are displayed by gray dashed lines. }}
	\label{fig:spin purification}
\end{figure}
To overcome such severe spin contamination, a common approach in quantum chemistry is to incorporate various penalty terms \cite{IPATOV200960, Li2024, FermiSci} into the loss function to widen the energy separation between nearly degenerate spin states. Nonetheless, this approach does not definitively ensure that the resulting state is the exact target spin eigenstate. In this work, we propose a novel spin purification scheme that instead of modifying the energy landscape, we directly remove all unwanted spin states, thereby ensuring more stable and accurate simulations.
This is achieved by applying spin projection operator $R_{s}$:
\begin{equation}
	\begin{aligned}
		R_{s} = \prod_{s_{i}>s} \left( \bm{S}^2 - s_{i}(s_{i} + 1) \right)
		\label{eq:spin-purification}
	\end{aligned}
\end{equation}
to the NQS, where $s_{i}$ labels the spin states to be removed. 
%$\bm{\sigma}_n$ is the Pauli matrix acting on baryon $n$
%In principle, one can construct a pure spin state by eliminating all unwanted spin configurations, although each application of 
%$R_{s_r}$
%increases the computational cost due to the multiplication of determinant counts by $C_A^2 + 1$. To save computational time, we selectively remove those primary contributors to spin contamination, which are typically spin states with $s_r = s+1$. 

Owing to the simplicity of the spectrum of hypernuclei, an consistent procedure can be used to treat both ground and excited states. To ensure that our ansatz remains orthogonal to all possible states with spin values less than $s$, we assign the spin magnetic quantum number $s_z=s$. A SR based minimization of energy is performed to determine the variational parameters in the spin purified NQS, $R_{S}\Psi^P$.

We here provide compact representations of the spin purification operators employed in our calculations. For the nuclei examined in this study, we have at most two unwanted states to remove. To begin with, if only an adjacent spin state is to be removed, $R_s$ can be expressed as
\begin{equation}
	\begin{aligned}
		R_{s} &=  \bm{S}^2 - s_{1}(s_{1} + 1) \\
			  &=  \frac{1}{4} \sum_{nm} \sigma_n \cdot \sigma_m - 	s_{1}(s_{1} + 1)\\
			  &= \sum_{n<m} P_{nm} + A - \frac{1}{4}A^2 - s_{1}(s_{1} + 1),
		\label{eq:single}
	\end{aligned}
\end{equation}
where $\bm{\sigma}_n$ is the Pauli matrix acting on baryon $n$ and $P_{nm}$ swaps the spins of the baryon pair $nm$. For the ground state of $^4_\Lambda$H, with two redundant spin states present, $R_s$ is structured as follows
\begin{equation}
	\begin{aligned}
		R_{s} &=  (\bm{S}^2 - s_{2}(s_{2} + 1))(\bm{S}^2 - s_{1}(s_{1} + 1)) \\
&= 2P_{12}P_{34} + 2P_{13}P_{24} + 2P_{14}P_{23} + 3 \sum_{n<m<l} \sum_{\text{cyc}} P_{nml} \\
&+ \left[ 2\left( A - \frac{1}{4}A^2 \right) - s_1(s_1 + 1) - s_2(s_2 + 1) \right] \sum_{n<m} P_{ij} \\
&+ \left( A - \frac{1}{4}A^2 \right)^2 - \left[ s_1(s_1 + 1)s_2(s_2 + 2) \right] \left( A - \frac{1}{4}A^2 \right)\\ &+ s_1(s_1 + 1)s_2(s_2 + 1) + C_A^2,
	\end{aligned}
\end{equation}
with $P_{nml}$ being the cyclic permutation over the triplet $nml$. Note that since all constant terms have been factored out, for each triplet, only two
permutations are considered, excluding the identity operator. 

Fig. \ref{fig:spin purification} aims to demonstrate the effectiveness of our spin enforcement approach, with the loss curve obtained with our method compared to that from the penalty method. We choose the penalty term to be proportional to spin square\cite{FermiSci} and modify the Hamiltonian 
\begin{align}
	H \rightarrow H + \omega_s \bm{S}^2,
\end{align}
where $\omega_s$ is set to 1 MeV and 2 MeV. In comparison with standard energy optimization, all spin purification methods markedly accelerate convergence process. 
Compared with penalty method, our spin purification approach enforces the NQS to be spin eigenstates, inherently eliminating potential spin contamination and thereby further enhancing both efficiency and accuracy. While our treatment is more computationally expensive than the penalty method, this is entirely acceptable for light hypernuclear systems. For heavier nuclei, however, the penalty method remains our preferred choice for spin purification, especially when spin-orbit and tensor forces are included. Nevertheless, incorporating spin symmetry constraints in nuclear VMC-NQS calculations proves necessary, particularly for nuclei with multiple degenerate spin states. 

\subsection*{Neural-network quantum states}
\begin{figure*}[tb]
	\centering
	\includegraphics[ width=\textwidth]{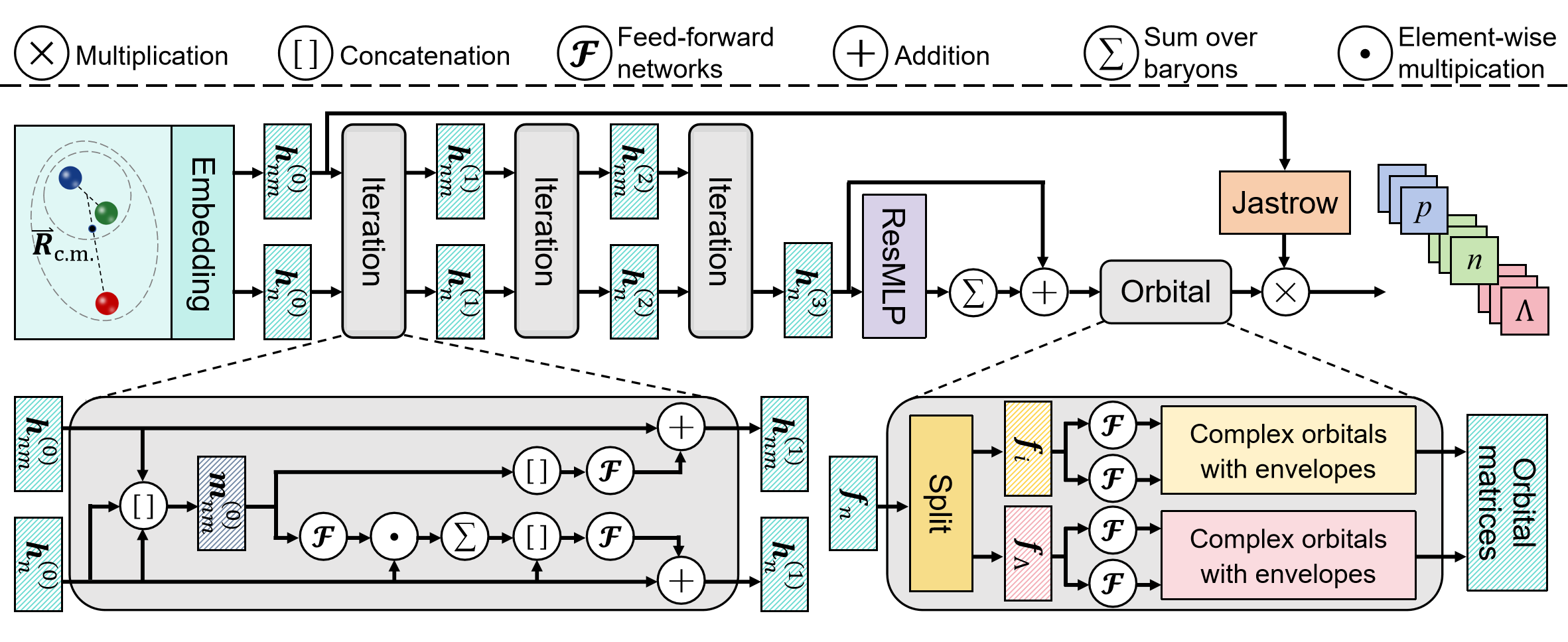}
	\caption{\parbox[t]{\textwidth}{FIG. 7. The neural-network quantum states for single-$\Lambda$ hypernuclei within SG formalism. Initial hidden features of baryons are collected and undergo three intermediate layers to form the final set of one-baryon features. These features are thereby refined by ResMLP and split into nucleon and hyperon features. Finally, these one-body features are used to construct three complex orbital matrices, which are further augmented by the Jastrow factor.}}
	\label{fig:network}
\end{figure*}
NQS are trainable mappings from many-body eigenstates to complex amplitudes, with essential symmetries imposed to reduce the redundant Hilbert space. Following the traditional NQS framework, the $\Lambda$ states can be incorporated by extending the input of the ansatz. For single-$\Lambda$ hypernuclei with mass number $A$, the NQS is a function $\Psi$: ($\bm{r}_i,s_i,t_i$)$^{A-1}\times(\bm{r}_\Lambda,s_\Lambda)\mapsto\mathbb{C}$, where $\bm{r}_{i(\Lambda)}\in\mathbb{R}^3$ and $s_{i(\Lambda)}\in\{1, -1\}$ denotes single-baryon spatial and spin degrees of freedom respectively. $t_i\in\{1, -1\}$ indicates the z-projection of isospins of the $A-1$ nucleons.
The SG procedure eliminates the intrinsic isospin degree of freedom of nucleons, allowing us to treat nucleons and $\Lambda$ in a unified manner. Thus the NQS reduces to $\Psi$: ($\bm{r}_n,s_n$)$^{A}\mapsto\mathbb{C}$. 

The starting phase of NQS is to construct initial one-baryon and two-baryon hidden features, which are chosen to be 
\begin{equation}
	\begin{aligned}
		&\bm{h}^{(0)}_{n} = \left( \bm{r}_{n}, \left\| \bm{r}_{n} \right\|, s_{n}\right), \\
		&\bm{h}^{(0)}_{nm} = \left( \bm{r}_{nm}, r_{nm}, s_{n}, s_m, s_{nm} \right), 
		\label{input}
	\end{aligned}
\end{equation}
with isospins grouped, where $\bm{r}_{nm}=\bm{r}_{m}-\bm{r}_{n}$, $r_{nm}=\left\| \bm{r}_{nm} \right\|$ and $s_{nm}=2\delta_{s_{n}s_m}-1$, capturing relative orientation of spins  of two baryons \cite{Kim2024}.
 Note that, NQS is ensured to be translational invariant by employing intrinsic spatial coordinates, defined by $\bm{r}_{n} \rightarrow \bm{r}_{n} - \bm{R}_{\text{c.m.}}$, with $\bm{R}_{\text{c.m.}}$ being the center of mass position
\begin{equation}
	\begin{aligned}
	\bm{R}_{\text{c.m.}} = \frac{M_N \sum_{i=1}^{A-1} \bm{r}_i + M_{\Lambda} \bm{r}_{\Lambda}}{M_N (A - 1) + M_{\Lambda}}.
	\end{aligned}
\end{equation}
The individual mass for nucleon and hyperon are $M_N=938.92$ MeV and $M_\Lambda=1115.68$ MeV. To make the sizes of the position and spin input comparable, we further perform a logarithm scaling \cite{vonglehn2023selfattentionansatzabinitioquantum, Laplacian} on the spatial part in Eq. (\ref{input}), which is given by $r \rightarrow \log \left( \frac{1 + |r |}{|r|} \right)\times r$.

Thereafter, correlations are built iteratively into the final hidden features within the MPNN framework \cite{MPNN}. In each intermediate layer $l$, the aggregation of relevant information begins with message formation\footnote{$\sigma$ is the activation function ($tanh$). All undefined variables in Eq. \eqref{eq:message formation}, \eqref{eq:sch-convlution}, \eqref{eq:feature updating}, \eqref{eq:envelope}, \eqref{eq:Jastrow} serve as variational parameters.}
\begin{equation}
	\begin{aligned}
		\bm{m}_{nm}^{(l)} = \sigma\left(\bm{W}^{(l)}\cdot\left( \bm{h}_n^{(l-1)}, \bm{h}_m^{(l-1)}, \bm{h}_{nm}^{(l-1)} \right) + \bm{w}^{(l)}\right). \label{eq:message formation}
	\end{aligned}
\end{equation}
Continuous kernels $\bm{K}^{(l)}_{nm}$ are generated afterwards based on pairwise messages and subsequently used to 
conduct SchNet-like convolution \cite{schnet}
\begin{equation}
	\begin{aligned}
		&\bm{K}^{(l)}_{nm} = \sigma\left(\bm{B}^{(l)}\cdot\bm{m}_{nm}^{(l)}+\bm{b}^{(l)}\right),\\
		&\bm{g}^{(l)}_{n} = \sum_m \bm{h}^{(l)}_m \odot \bm{K}^{(l)}_{nm}, \label{eq:sch-convlution}
	\end{aligned}
\end{equation}
where \enquote{$\odot$} denotes element-wise multiplication. $\bm{g}^{(l)}_n$ and $\bm{m}^{(l)}_{nm}$ are then used to update one- and two-baryon hidden features in a similar manner
\begin{equation}
	\begin{aligned}
		&\bm{h}^{(l+1)}_{n} = \sigma\left(\bm{U}^{(l)}\cdot\left(\bm{h}^{(l)}_n, \bm{g}^{(l)}_n\right) + \bm{u}^{(l)}\right)+\bm{h}^{(l)}_n,\\
		&\bm{h}^{(l+1)}_{nm} = \sigma\left(\bm{V}^{(l)}\cdot\left(\bm{h}^{(l)}_{nm}, \bm{m}^{(l)}_{nm}\right) + \bm{v}^{(l)}\right)+\bm{h}^{(l)}_{nm}.
	\end{aligned}  \label{eq:feature updating}
\end{equation}
We add shortcut branches to address potential vanishing or exploding gradients \cite{resnet}. After our last intermediate layer, labeled as $L$, one-baryon features are collected and further refined exploiting multilayer perceptrons with residual connections in each layer (ResMLP)
\begin{equation}
	\begin{aligned}
		\left(\bm{f}_i, \bm{f}_\Lambda\right)=\bm{f}_n = \bm{h}^{(L)}_n + \sum_m \mathrm{ResMLP}\left(\bm{h}^{(L)}_m\right).
	\end{aligned}  
\end{equation}
Finally, the backflow transformed one-body features, $\bm{f}_i$ and $\bm{f}_\Lambda$, are constructed with sufficient capability to capture baryonic correlations.

The trial states are practically constructed in a Slater-Jastrow form
\begin{equation}
	\begin{aligned}
		&\langle\mathbf{x}|\Psi\rangle = e^{J\left(\mathbf{x}\right)}\times\\
		&\sum_{k=1}^K\det\left[\phi^{(k)}_p(\mathbf{x})\right]\det\left[\phi^{(k)}_n(\mathbf{x})\right]\det\left[\phi^{(k)}_\Lambda(\mathbf{x})\right],
	\end{aligned}
\end{equation}
where $\mathbf{x}$ refers to a set of baryonic position-spin eigenstates. To construct $\phi^{(k)}(\mathbf{x})$, we first feed $\bm{f}_{i(\Lambda)}$ into two distinct feed-forward networks, forming the real and imaginary components of orbitals. These orbitals are subsequently multiplied with learnable envelopes to suppress fragmentation into sub-clusters
\begin{equation}
	\begin{aligned}
		&\mathrm{env}^k_i\left(\bm{r}_j\right)=\exp\left(-|\xi^k_i|\times\|\bm{r}_j-\bm{R}_{\mathrm{n.c.}}\|\right),\\
		&\mathrm{env}^k_\Lambda\left(\bm{r}_{\Lambda }\right)=\exp\left(-\|\bm{\eta}^k_i\odot(\bm{r}_\Lambda-\bm{R}_{\mathrm{n.c.}})\|\right), \label{eq:envelope}
	\end{aligned}
\end{equation}
where $\bm{R}_{\mathrm{n.c.}}$ represents the coordinate of nuclear core. For nucleons, a vanilla isotropic exponential decay is used to confine them around $\bm{R}_{\mathrm{n.c.}}$. To model the complicated tail of $\Lambda$ distribution, this work adopts the anisotropic envelope close to FermiNet \cite{Ferminet}. The Jastrow factor $J\left(\mathbf{x}\right)$  reads
\begin{equation}
	\begin{aligned}
		J\left(\mathbf{x}\right) = \sigma\biggl( \boldsymbol{G} \cdot \biggl( & \sum_{i\neq j}\mathrm{ResMLP}\left(\boldsymbol{h}^{(0)}_{ij}\right), \\
		& \sum_{j}\mathrm{ResMLP}\left(\boldsymbol{h}^{(0)}_{\Lambda j}\right) \biggr) + \boldsymbol{g} \biggr). \label{eq:Jastrow}
	\end{aligned}
\end{equation}
%Since NQS inherently captures Kato cusps, the role of Jastrow factor shifts to eliminating explicit product-state artifacts, thereby obviating the need to retain the traditional exponential form.
Incorporating an additional correlation factor can eliminate explicit product-state artifacts, thereby enhancing the expressibility of NQS with limited configurations. We provide a graphical visualization of our ansatz architecture in Fig. \ref{fig:network}. 

For the NQS without SG technique, the primary distinction lies in the embedding layer, as a result of the misalignment in particle degrees of freedom. We take initial features as
\begin{equation}
	\begin{aligned}
		&\bm{h}^{(0)}_{i} = \left( \bm{r}_{i}, \left\| \bm{r}_{i} \right\|, s_{i}, t_{i} \right), \\
		&\bm{h}^{(0)}_{\Lambda} = \left( \bm{r}_{\Lambda}, \left\| \bm{r}_{\Lambda} \right\|, s_{i}\right), \\
		&\bm{h}^{(0)}_{ij} = \left( \bm{r}_{ij}, r_{ij}, s_{i}, s_j, s_{ij}, t_{i}, t_j, t_{ij} \right), \\
		&\bm{h}^{(0)}_{\Lambda j} = \left( \bm{r}_{\Lambda j}, r_{\Lambda j}, s_{\Lambda}, s_j, s_{\Lambda j} \right).
	\end{aligned}
\end{equation} 
The dimensions of one- and two-body features are following aligned by trainable linear transformations with bias, expressed as 
$\bm{h}^{(0)}_{\alpha} \rightarrow \bm{A}_{\alpha} \cdot \bm{h}^{(0)}_{\alpha} + \bm{a}_\alpha$,
where \( \alpha \) represents one of the indices \(\{i, \Lambda, ij, \Lambda j\}\).
The primary objective of dimension alignment is to establish the initial hidden one- and two-baryon features \( \bm{h}^{(0)}_n \) and \( \bm{h}^{(0)}_{nm} \). This is achieved through the concatenation of \( \bm{h}^{(0)}_i \) and \( \bm{h}^{(0)}_\Lambda \), as well as \( \bm{h}^{(0)}_{ij} \) and \( \bm{h}^{(0)}_{\Lambda j} \), along the first axis. We use the identical MPNN, Jastrow factor and orbital forming block as that in SG version. However, we cannot block diagonalize the nucleon determinant and thus must compute its full form.

%Guided by Rayleigh-Ritz variational principle, the expectation of Hamiltonian is evaluated via Metropolis-Hastings sampling algorithm detailed in the  supplemental materials of Ref. \cite{v2007} and minimized iteratively employing RMSprop-regularized stochastic
%reconfiguration optimizer \cite{Hidden, Feynmannet}. Furthermore, to stabilize training process, we design Glide SR algorithm for learning rate scheduling. 
\subsection*{Sampling}
On the suggestion of S.D. Chai (personal communication), instead of MH sampling introduced in the supplementary material of Ref~\cite{v2007}, we design a Gibbs-inspired MH sampling (MH-Gibbs) in our hypernuclear calculations. The key improvement is instead of directly generating samples from the joint distribution $|\Psi(\bm{r}_i, \bm{r}_\Lambda, \chi_n)|^2$, we sample on the conditional distribution of each individual degree of freedom sequentially. $\chi_n$ refers to $(s_i, t_i)$ for nucleons and $s_\Lambda$ for $\Lambda$ hyperons. This considerably reduces the dimensionality of the sampling space, thereby making the detailed balance condition easier to satisfy. The separation of degrees of freedom has been empirically recommended for MH sampling in nuclear VMC-NQS calculations\cite{v2007}, despite the lack of theoretical backing. Gibbs inspired strategy offers a more theoretically sound and scalable alternative. 

The pseudocode of MH-Gibbs sampling is shown in Alg. \ref{alg: Gibbs sampling}. For hypernuclei, we treat the nucleon positions, hyperon positions and spinors as three sets of 
variables. Between two optimization steps, samples are propagated through $T=10$ iterations. 
\begin{figure}[tbp]
	\includegraphics[width=0.49\textwidth]{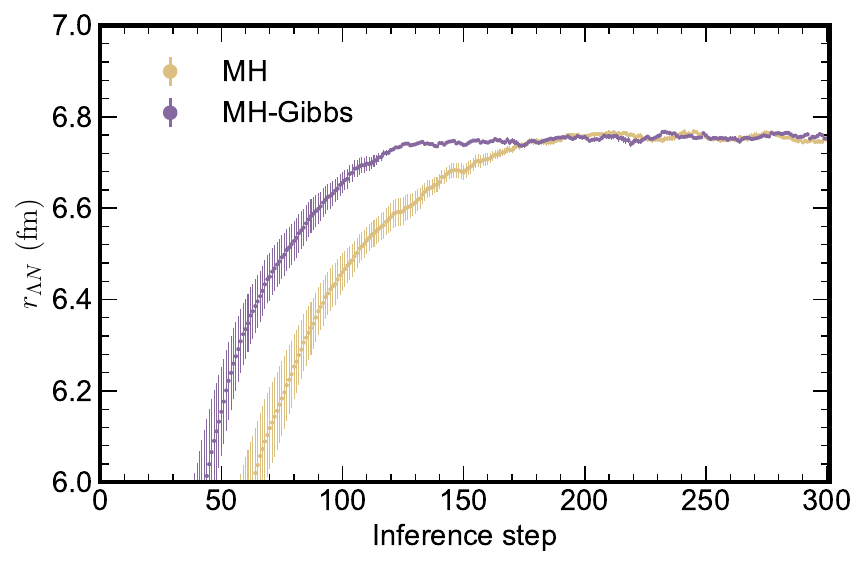}
    	\caption{\parbox[t]{0.49\textwidth}{FIG. 8. We carry out separate inferences from a $^{13}_\Lambda$C NQS with MH and MH-Gibbs algorithm. The distance between $\Lambda$ and the $^{12}$C core is plotted as a function of inference step. In each step, we run a 100-step MH sampling. The times of wavefunction calculations are identical for both algorithms.}}
	\label{fig:sampling}
\end{figure}
\begin{algorithm}[t]
	\caption{MH-Gibbs sampling}
	\label{alg: Gibbs sampling}
	\begin{algorithmic}[1]
		\Require{initial samples $\{(\bm{r}^{(0)}_i, \bm{r}^{(0)}_\Lambda,\chi^{(0)}_n)\}$, iterations $T$} 
		\For{$t$ in \{1, ..., $T$\}}
		\State {$\bm{r}^{(t)}_i\sim |\Psi(\bm{r}_i, \bm{r}_\Lambda=\bm{r}^{(t-1)}_\Lambda, \chi_n=\chi^{(t-1)}_n)|^2$}
		\State {$\bm{r}^{(t)}_\Lambda\sim |\Psi(\bm{r}_i=\bm{r}^{(t)}_i, \bm{r}_\Lambda, \chi_n=\chi^{(t-1)}_n)|^2$}
		\State {$\chi^{(t)}_n\sim |\Psi(\bm{r}_i=\bm{r}^{(t)}_i, \bm{r}_\Lambda=\bm{r}^{(t)}_\Lambda, \chi_n|^2$}
	\EndFor\\
		\Return $\{(\bm{r}^{(T)}_i, \bm{r}^{(T)}_\Lambda,\chi^{(T)}_n)\}$
	\end{algorithmic}
\end{algorithm}
In each iteration, we sample the three degrees of freedom from the marginal distribution subsequently, with a 10-step MH process. Noting that this MH-Gibbs sampling can be considered as a special case of standard MH sampling, with each proposal move only kicking one degree of freedom. Our MH-Gibbs algorithm shares commons with the shuttle algorithm for the sampling of auxiliary field configurations in NLEFT calculations, in which only one time slice is updated at a time \cite{Lu:2018bat}. Performance of MH-Gibbs sampling has been investigated by comparing the inference speed of a $^{13}_\Lambda$C wavefunction to MH sampling in Fig. \ref{fig:sampling}. Compared to MH sampling, MH-Gibbs algorithm takes fewer inference steps to achieve detailed balance, about 100 steps. We expect MH-Gibbs sampling to be particularly beneficial for heavier systems and those with a higher particle diversity.

\subsection*{Optimization}
\begin{figure}[tbp]
	\includegraphics[width=0.49\textwidth]{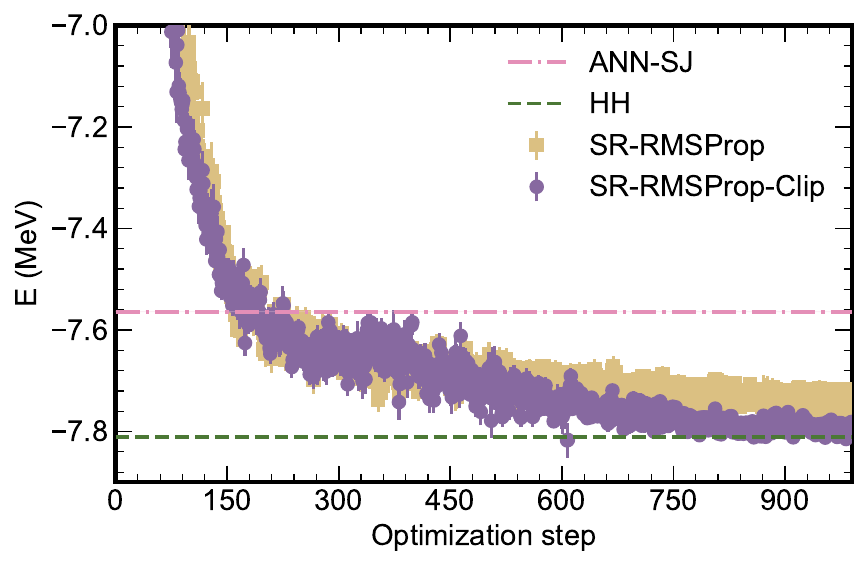}
	\caption{\parbox[t]{0.49\textwidth}{FIG. 9. We present the convergence parttern for $^3$He using SR-RMSProp (yellow solid square) and SR-RMSProp-Clip (purple solid circle) algorithm, with identical hyperparameter set up. For clipped version, we set $d_{FS}=0.02$, $\gamma=2\times10^{-4}$, $\Gamma=1500$. In SR-RMSProp algorithm, we keep the learning rate fixed at $2\times10^{-4}$
		throughout the training. Results from the ANN-SJ ansatz and
			HH method \cite{A6} are shown as pink dash-dotted lines and green dashed lines, with shaded area as the error bars. }}
	\label{fig:schedule}
\end{figure}
In this subsection, we discuss the wisdom of optimization in VMC-NQS calculation. The basic principle of minimizing a given loss function $L(\bm{p}_t)$ within gradient decent framework is the variational trial of parameters $\bm{p}_t$, with the norm of deviation $\delta \bm{p}_t$ constrained
\begin{equation}
	\left\{
	\begin{aligned}
		&\delta \bm{p}_t = \arg \min L(\bm{p}_t + \delta\bm{p}_t), \\
		&\|\delta\bm{p}_t\|^2 = \delta\bm{p}_t^T\bm{G}_t\delta\bm{p}_t=c_t.
	\end{aligned}
	\right.
	\label{minimize}
\end{equation}
$\bm{G}_t$ is the metric of parameter space. In the first order approximation, Lagrange multiplier method gives the optimal direction of each iteration of parameter search is proportional to $-\bm{G}_t^{-1}\bm{g}_t$. $\bm{g}_t$ is the gradient of loss function in $t$ step. The landscape of parameter space is characterized by the inverse of metric. For quantum states living in complex projective space, the gauge symmetry can be respected naturally by defining Fubini-Study (FS) distance \cite{Stokes2020}
\begin{equation}
	d\left(\Phi, \Psi\right)=\arccos\left(\sqrt{\dfrac{\langle\Phi|\Psi\rangle\langle\Psi|\Phi\rangle}{\langle\Phi|\Phi\rangle\langle\Psi|\Psi\rangle}}\right).
\end{equation}
One can show the corresponding metric reads
\begin{equation}
	\begin{aligned}
		\bm{G}_t &= \mathfrak{R}\left[\langle\partial_{\bm{p}_t}^T\partial_{\bm{p}_t}\rangle-\langle\partial_{\bm{p}_t}^T\rangle\langle\partial_{\bm{p}_t}\rangle\right],\\
	\end{aligned}
\end{equation}
defined as the real part of quantum geometric tensor (QGT). $\langle O \rangle$ denotes the expectation of a certain operator in state $|\Psi\rangle$, which is computed by meaning $O\Psi/\Psi$ over a large set of samples generated following the distribution of $|\Psi|^2$, denoted as $\mathbb{E}_{\Psi}\left[O\Psi/\Psi\right]$. In practical implementation, we express $\Psi$ as $\rho\exp(i\theta)$, and have
\begin{equation}
	\begin{aligned}
		\dfrac{\partial_{\bm{p}_t}\Psi}{\Psi}=\partial_{\bm{p}_t}\ln\rho +  i \partial_{\bm{p}_t}\theta.
	\end{aligned}
	\label{logsign}
\end{equation}
Such that the FS metric can be computed as
\begin{align}
\bm{G}_t &= \mathbb{E}_{\Psi}\left[\left(\partial_{\bm{p}_t}\ln\rho - \mathbb{E}_{\Psi}\left[\partial_{\bm{p}_t}\ln\rho\right]\right)^T\left(\partial_{\bm{p}_t}\ln\rho - \mathbb{E}_{\Psi}\left[\partial_{\bm{p}_t}\ln\rho\right]\right)\right] \nonumber \\
&+ \mathbb{E}_{\Psi}\left[\left(\partial_{\bm{p}_t}\theta - \mathbb{E}_{\Psi}\left[\partial_{\bm{p}_t}\theta\right]\right)^T\left(\partial_{\bm{p}_t}\theta - \mathbb{E}_{\Psi}\left[\partial_{\bm{p}_t}\theta\right]\right)\right].
\end{align}
During the realization, the inversion of large-scale matrix is avoided by instead solving a linear system with $\bm{G}_t$ as the coefficient matrix. We here employ the Cholesky decomposition, with $\bm{G}_t$ ensured to be positive definite via RMSProp regularization\cite{Hidden, Feynmannet}.

The idealized gradient descent aims to follow a smooth trajectory within the parameter space. This can be materialized by taking a succession of small steps, each characterized by a small learning rate $\alpha_t$. However, poor parameter initialization can lead to erratic gradient variations, hindering smooth NQS changes. Following Eq. (\ref{minimize}), one can write down the following constraint
\begin{equation}
	\alpha_t^2 (\bm{G}_t^{-1} \bm{g}_t)^T \bm{G}_t \bm{G}_t^{-1} \bm{g}_t = \alpha_t^2 \bm{g}_t^T \bm{G}_t^{-1} \bm{g}_t \leq d_{FS}^2,
	\label{constraint}
\end{equation}
which suggests the variation of quantum state remains below $d_{FS}$ in each step. Such constraint can be imposed by setting the learning rate to be
\begin{equation}
	\alpha_t = \text{min}\left(\dfrac{d_{FS}}{\sqrt{\bm{g}_t^T \bm{G}_t^{-1} \bm{g}_t}}, \gamma\right)\times\frac{1}{1 + t/\Gamma},
	\label{clipping}
\end{equation}
which is close to the form of gradient clipping. We further multiply the learning rate with a decay factor to perform more meticulous parameter adjustment near the optimal point.

In Fig. \ref{fig:schedule}, we compare the performance of the original SR-RMSProp minimizer to the version that employs the learning rate defined in Eq. (\ref{clipping}), which we dub SR-RMSProp-Clip. It is evident that our clipped version can not only achieve smaller variance, but also yield a ground energy noticeably closer to the numerically exact HH result. This indicates SR-RMSProp-Clip method can find a better minimum than the SR-RMSProp.

\section*{Acknowledgments}
We are grateful for fruitful discussions with Sheng-Du Chai, Ling-Yi Dai, Serdar Elhatisari, Bo Huang, Jane Kim, Guo-Ping Li, Pei Li, Xin Li, Dai-Neng Liu, Zhao-Xi Shen, Kai-Jia Sun, Si-Min Wang, Zhi-Cheng Xu, Jin-Cheng Yang and Bo Zhou. We would like to thank Martin Schäfer for providing  the LECs of the hyperon-nucleon interactions and the SVM results on $s$-shell hypernuclei sourced in Ref.~\cite{PhysRevC.106.L031001}. This work is partially supported by the National Natural Science Foundation of China under Contracts No. 12147101,  12475117, 12141501 and 123B2080, the Guangdong Major Project of Basic and Applied Basic Research No. 2020B0301030008, and the STCSM under Grant No. 23590780100 and 23JC1400200, the National Key Research and Development Project of China under Grant No. 2022YFA1604900, 2024YFA1610702 and 2024YFA1612600. The computations in this research were performed using the CFFF platform of Fudan University. This work has been also supported by the High-performance Computing Platform of Peking University.

\section*{Author Contributions}
Zi-Xiao Zhang designed the theoretical framework, developed the code, and performed the numerical simulations. Wan-Bing He provided overall supervision and guidance throughout the project. Yi-Long Yang and Peng-Wei Zhao provided their results on atomic nuclei and hypernuclei for benchmarking, calculated with FeynmanNet. Bing-Nan Lu taught Zi-Xiao Zhang the details of NLEFT caculations. Yu-Gang Ma offered valuable insights through critical reviews of the manuscript.
Zi-Xiao Zhang extends gratitude for the enlightening discussions with all coauthors who have significantly contributed to writing the manuscript.
\bibliographystyle{wlscirep}
\bibliography{references}

\begin{thebibliography}{10}
\urlstyle{rm}
\expandafter\ifx\csname url\endcsname\relax
  \def\url#1{\texttt{#1}}\fi
\expandafter\ifx\csname urlprefix\endcsname\relax\def\urlprefix{URL }\fi
\expandafter\ifx\csname doiprefix\endcsname\relax\def\doiprefix{DOI: }\fi
\providecommand{\bibinfo}[2]{#2}
\providecommand{\eprint}[2][]{\url{#2}}

\bibitem{RevModPhys.88.035004}
\bibinfo{author}{Gal, A.}, \bibinfo{author}{Hungerford, E.~V.} \& \bibinfo{author}{Millener, D.~J.}
\newblock \bibinfo{journal}{\bibinfo{title}{Strangeness in nuclear physics}}.
\newblock {\emph{\JournalTitle{Rev. Mod. Phys.}}} \textbf{\bibinfo{volume}{88}}, \bibinfo{pages}{035004}, \doiprefix\url{10.1103/RevModPhys.88.035004} (\bibinfo{year}{2016}).

\bibitem{Yu-Gang2017}
\bibinfo{author}{Ma, Y.-G.}, \bibinfo{author}{Chen, J.-H.}, \bibinfo{author}{Liu, P.} \& \bibinfo{author}{Zhang, S.}
\newblock \bibinfo{journal}{\bibinfo{title}{Production of light nuclei and hypernuclei at high intensity accelerator facility energy region}}.
\newblock {\emph{\JournalTitle{Nuclear Science and Techniques}}} \textbf{\bibinfo{volume}{28}}, \bibinfo{pages}{55}, \doiprefix\url{10.1007/s41365-017-0207-x} (\bibinfo{year}{2017}).

\bibitem{Chen:2023mel}
\bibinfo{author}{Chen, J.}, \bibinfo{author}{Dong, X.}, \bibinfo{author}{Ma, Y.-G.} \& \bibinfo{author}{Xu, Z.}
\newblock \bibinfo{journal}{\bibinfo{title}{{Measurements of the lightest hypernucleus (H{\ensuremath{\Lambda}}3): progress and perspective}}}.
\newblock {\emph{\JournalTitle{Sci. Bull.}}} \textbf{\bibinfo{volume}{68}}, \bibinfo{pages}{3252--3260}, \doiprefix\url{10.1016/j.scib.2023.11.045} (\bibinfo{year}{2023}).
\newblock \eprint{2311.09877}.

\bibitem{PhysRevLett.134.022301}
\bibinfo{author}{Sun, K.-J.} \emph{et~al.}
\newblock \bibinfo{journal}{\bibinfo{title}{Deciphering hypertriton and antihypertriton spins from their global polarizations in heavy-ion collisions}}.
\newblock {\emph{\JournalTitle{Phys. Rev. Lett.}}} \textbf{\bibinfo{volume}{134}}, \bibinfo{pages}{022301}, \doiprefix\url{10.1103/PhysRevLett.134.022301} (\bibinfo{year}{2025}).

\bibitem{STAR2010}
\bibinfo{author}{{The STAR Collaboration}}.
\newblock \bibinfo{journal}{\bibinfo{title}{Observation of an antimatter hypernucleus}}.
\newblock {\emph{\JournalTitle{Science}}} \textbf{\bibinfo{volume}{328}}, \bibinfo{pages}{58--62}, \doiprefix\url{10.1126/science.1183980} (\bibinfo{year}{2010}).

\bibitem{STAR2024}
\bibinfo{author}{{The STAR Collaboration}}.
\newblock \bibinfo{journal}{\bibinfo{title}{Observation of the antimatter hypernucleus \(_{\bar{\Lambda}}^4 \bar{{\rm{H}}}\)}}.
\newblock {\emph{\JournalTitle{Nature}}} \textbf{\bibinfo{volume}{632}}, \bibinfo{pages}{1026--1031}, \doiprefix\url{10.1038/s41586-024-07823-0} (\bibinfo{year}{2024}).

\bibitem{PhysRevLett.134.162301}
\bibinfo{author}{{The ALICE Collaboration}}.
\newblock \bibinfo{journal}{\bibinfo{title}{First measurement of $a=4$ hypernuclei and antihypernuclei at the lhc}}.
\newblock {\emph{\JournalTitle{Phys. Rev. Lett.}}} \textbf{\bibinfo{volume}{134}}, \bibinfo{pages}{162301}, \doiprefix\url{10.1103/PhysRevLett.134.162301} (\bibinfo{year}{2025}).

\bibitem{Chen:2018tnh}
\bibinfo{author}{Chen, J.}, \bibinfo{author}{Keane, D.}, \bibinfo{author}{Ma, Y.-G.}, \bibinfo{author}{Tang, A.} \& \bibinfo{author}{Xu, Z.}
\newblock \bibinfo{journal}{\bibinfo{title}{{Antinuclei in Heavy-Ion Collisions}}}.
\newblock {\emph{\JournalTitle{Phys. Rept.}}} \textbf{\bibinfo{volume}{760}}, \bibinfo{pages}{1--39}, \doiprefix\url{10.1016/j.physrep.2018.07.002} (\bibinfo{year}{2018}).
\newblock \eprint{1808.09619}.

\bibitem{Chen:2024aom}
\bibinfo{author}{Chen, J.} \emph{et~al.}
\newblock \bibinfo{journal}{\bibinfo{title}{{Properties of the QCD matter: review of selected results from the relativistic heavy ion collider beam energy scan (RHIC BES) program}}}.
\newblock {\emph{\JournalTitle{Nucl. Sci. Tech.}}} \textbf{\bibinfo{volume}{35}}, \bibinfo{pages}{214}, \doiprefix\url{10.1007/s41365-024-01591-2} (\bibinfo{year}{2024}).
\newblock \eprint{2407.02935}.

\bibitem{2025halo}
\bibinfo{author}{Knöll, M.} \& \bibinfo{author}{Roth, R.}
\newblock \bibinfo{title}{Halo structures in p-shell hypernuclei with natural orbitals} (\bibinfo{year}{2025}).
\newblock \eprint{2501.08013}.

\bibitem{Yu-GangMa2023}
\bibinfo{author}{Ma, Y.-G.}
\newblock \bibinfo{journal}{\bibinfo{title}{Hypernuclei as a laboratory to test hyperon-nucleon interactions}}.
\newblock {\emph{\JournalTitle{Nuclear Science and Techniques}}} \textbf{\bibinfo{volume}{34}}, \bibinfo{pages}{97}, \doiprefix\url{10.1007/s41365-023-01248-6} (\bibinfo{year}{2023}).

\bibitem{LIU2024138855}
\bibinfo{author}{Liu, D.-N.} \emph{et~al.}
\newblock \bibinfo{journal}{\bibinfo{title}{Softening of the hypertriton transverse momentum spectrum in heavy-ion collisions}}.
\newblock {\emph{\JournalTitle{Physics Letters B}}} \textbf{\bibinfo{volume}{855}}, \bibinfo{pages}{138855}, \doiprefix\url{https://doi.org/10.1016/j.physletb.2024.138855} (\bibinfo{year}{2024}).

\bibitem{Chen2025}
\bibinfo{author}{Chen, J.-H.} \emph{et~al.}
\newblock \bibinfo{journal}{\bibinfo{title}{Production of exotic hadrons in pp and nuclear collisions}}.
\newblock {\emph{\JournalTitle{Nuclear Science and Techniques}}} \textbf{\bibinfo{volume}{36}}, \bibinfo{pages}{55}, \doiprefix\url{10.1007/s41365-025-01664-w} (\bibinfo{year}{2025}).

\bibitem{Meoto_2020}
\bibinfo{author}{Meoto, E.~F.} \& \bibinfo{author}{Lekala, M.~L.}
\newblock \bibinfo{journal}{\bibinfo{title}{Faddeev calculations on lambda hypertriton with potentials from {G}el{'} fand{-}levitan{-}marchenko theory}}.
\newblock {\emph{\JournalTitle{Communications in Theoretical Physics}}} \textbf{\bibinfo{volume}{72}}, \bibinfo{pages}{105302}, \doiprefix\url{10.1088/1572-9494/aba25a} (\bibinfo{year}{2020}).

\bibitem{Faddev}
\bibinfo{author}{Nogga, A.}, \bibinfo{author}{Kamada, H.} \& \bibinfo{author}{Gl\"ockle, W.}
\newblock \bibinfo{journal}{\bibinfo{title}{The hypernuclei $_{\ensuremath{\Lambda}}^{4}${He} and $_{\ensuremath{\Lambda}}^{4}${H}: Challenges for modern hyperon-nucleon forces}}.
\newblock {\emph{\JournalTitle{Phys. Rev. Lett.}}} \textbf{\bibinfo{volume}{88}}, \bibinfo{pages}{172501}, \doiprefix\url{10.1103/PhysRevLett.88.172501} (\bibinfo{year}{2002}).

\bibitem{Hiyama_2012}
\bibinfo{author}{Hiyama, E.} \& \bibinfo{author}{Yamamoto, Y.}
\newblock \bibinfo{journal}{\bibinfo{title}{Structure of $^{10}${B}e and $^{10}${B} hypernuclei studied with the four-body cluster model}}.
\newblock {\emph{\JournalTitle{Progress of Theoretical Physics}}} \textbf{\bibinfo{volume}{128}}, \bibinfo{pages}{105124}, \doiprefix\url{10.1143/ptp.128.105} (\bibinfo{year}{2012}).

\bibitem{2024vmc}
\bibinfo{author}{Sharma, B.}
\newblock \bibinfo{title}{Variational study of light hypernuclei} (\bibinfo{year}{2024}).
\newblock \eprint{2411.04894}.

\bibitem{Usmani_2006}
\bibinfo{author}{Usmani, A.~A.}
\newblock \bibinfo{journal}{\bibinfo{title}{$\ensuremath{\Lambda}{N}$ space-exchange correlation effects in the ${}_{\ensuremath{\Lambda}}^{5}\text{He}$ hypernucleus}}.
\newblock {\emph{\JournalTitle{Phys. Rev. C}}} \textbf{\bibinfo{volume}{73}}, \bibinfo{pages}{011302}, \doiprefix\url{10.1103/PhysRevC.73.011302} (\bibinfo{year}{2006}).

\bibitem{Usmani_2008}
\bibinfo{author}{Usmani, A.~A.} \& \bibinfo{author}{Khanna, F.~C.}
\newblock \bibinfo{journal}{\bibinfo{title}{Behaviour of the ${\Lambda} {N}$ and ${\Lambda} {NN}$ potential strengths in the ${}^5_{\Lambda}${H}e hypernucleus}}.
\newblock {\emph{\JournalTitle{Journal of Physics G: Nuclear and Particle Physics}}} \textbf{\bibinfo{volume}{35}}, \bibinfo{pages}{025105}, \doiprefix\url{10.1088/0954-3899/35/2/025105} (\bibinfo{year}{2008}).

\bibitem{LONARDONI2013243}
\bibinfo{author}{Lonardoni, D.}, \bibinfo{author}{Pederiva, F.} \& \bibinfo{author}{Gandolfi, S.}
\newblock \bibinfo{journal}{\bibinfo{title}{Auxiliary field diffusion monte carlo study of the hyperon–nucleon interaction in ${\Lambda}${-}hypernuclei}}.
\newblock {\emph{\JournalTitle{Nuclear Physics A}}} \textbf{\bibinfo{volume}{914}}, \bibinfo{pages}{243--247}, \doiprefix\url{https://doi.org/10.1016/j.nuclphysa.2012.12.001} (\bibinfo{year}{2013}).

\bibitem{PhysRevC89}
\bibinfo{author}{Lonardoni, D.}, \bibinfo{author}{Pederiva, F.} \& \bibinfo{author}{Gandolfi, S.}
\newblock \bibinfo{journal}{\bibinfo{title}{Accurate determination of the interaction between $\ensuremath{\Lambda}$ hyperons and nucleons from auxiliary field diffusion monte carlo calculations}}.
\newblock {\emph{\JournalTitle{Phys. Rev. C}}} \textbf{\bibinfo{volume}{89}}, \bibinfo{pages}{014314}, \doiprefix\url{10.1103/PhysRevC.89.014314} (\bibinfo{year}{2014}).

\bibitem{Frame2020ImpurityLM}
\bibinfo{author}{Frame, D.~K.}, \bibinfo{author}{L{\"a}hde, T.~A.}, \bibinfo{author}{Lee, D.} \& \bibinfo{author}{Meissner, U.-G.}
\newblock \bibinfo{journal}{\bibinfo{title}{Impurity lattice monte carlo for hypernuclei}}.
\newblock {\emph{\JournalTitle{The European Physical Journal A}}} \textbf{\bibinfo{volume}{56}} (\bibinfo{year}{2020}).

\bibitem{Hildenbrand2022}
\bibinfo{author}{Hildenbrand, F.}, \bibinfo{author}{Elhatisari, S.}, \bibinfo{author}{Lähde, T.~A.}, \bibinfo{author}{Lee, D.} \& \bibinfo{author}{Meißner, U.-G.}
\newblock \bibinfo{journal}{\bibinfo{title}{Lattice {Monte} {Carlo} simulations with two impurity worldlines}}.
\newblock {\emph{\JournalTitle{The European Physical Journal A}}} \textbf{\bibinfo{volume}{58}}, \bibinfo{pages}{167}, \doiprefix\url{10.1140/epja/s10050-022-00821-8} (\bibinfo{year}{2022}).

\bibitem{Hildenbrand2024}
\bibinfo{author}{Hildenbrand, F.}, \bibinfo{author}{Elhatisari, S.}, \bibinfo{author}{Ren, Z.} \& \bibinfo{author}{Meißner, U.~G.}
\newblock \bibinfo{journal}{\bibinfo{title}{Towards hypernuclei from nuclear lattice effective field theory}}.
\newblock {\emph{\JournalTitle{The European Physical Journal A}}} \textbf{\bibinfo{volume}{60}}, \bibinfo{pages}{215}, \doiprefix\url{10.1140/epja/s10050-024-01427-y} (\bibinfo{year}{2024}).

\bibitem{tong2025abinitiocalculationhyperneutron}
\bibinfo{author}{Tong, H.}, \bibinfo{author}{Elhatisari, S.} \& \bibinfo{author}{Meißner, U.-G.}
\newblock \bibinfo{title}{Ab initio calculation of hyper-neutron matter} (\bibinfo{year}{2025}).
\newblock \eprint{2405.01887}.

\bibitem{LI2025139708}
\bibinfo{author}{Li, X.}, \bibinfo{author}{Michel, N.}, \bibinfo{author}{Li, J.} \& \bibinfo{author}{Zhou, X.-R.}
\newblock \bibinfo{journal}{\bibinfo{title}{Gamow shell model description of neutron-rich he hyper-isotopes}}.
\newblock {\emph{\JournalTitle{Physics Letters B}}} \textbf{\bibinfo{volume}{868}}, \bibinfo{pages}{139708}, \doiprefix\url{https://doi.org/10.1016/j.physletb.2025.139708} (\bibinfo{year}{2025}).

\bibitem{Le2025}
\bibinfo{author}{Le, H.} \emph{et~al.}
\newblock \bibinfo{journal}{\bibinfo{title}{Benchmarking ${\Lambda}$${NN}$ three - body forces and first predictions for ${A} = 3 - 5$ hypernuclei}}.
\newblock {\emph{\JournalTitle{The European Physical Journal A}}} \textbf{\bibinfo{volume}{61}}, \bibinfo{pages}{21}, \doiprefix\url{10.1140/epja/s10050 - 024 - 01474 - 5} (\bibinfo{year}{2025}).

\bibitem{PRLLe}
\bibinfo{author}{Le, H.}, \bibinfo{author}{Haidenbauer, J.}, \bibinfo{author}{Mei\ss{}ner, U.-G.} \& \bibinfo{author}{Nogga, A.}
\newblock \bibinfo{journal}{\bibinfo{title}{Light $\mathrm{\ensuremath{\Lambda}}$ hypernuclei studied with chiral hyperon-nucleon and hyperon-nucleon-nucleon forces}}.
\newblock {\emph{\JournalTitle{Phys. Rev. Lett.}}} \textbf{\bibinfo{volume}{134}}, \bibinfo{pages}{072502}, \doiprefix\url{10.1103/PhysRevLett.134.072502} (\bibinfo{year}{2025}).

\bibitem{Nemura1999StudyOL}
\bibinfo{author}{Nemura, H.}, \bibinfo{author}{Suzuki, Y.}, \bibinfo{author}{Fujiwara, Y.} \& \bibinfo{author}{Nakamoto, C.}
\newblock \bibinfo{journal}{\bibinfo{title}{Study of light ${\Lambda}$- and ${\Lambda}$${\Lambda}$-hypernuclei with the stochastic variational method and effective ${\Lambda} {N}$ potentials}}.
\newblock {\emph{\JournalTitle{Progress of Theoretical Physics}}} \textbf{\bibinfo{volume}{103}}, \bibinfo{pages}{929--958} (\bibinfo{year}{1999}).

\bibitem{PhysRevC.106.L031001}
\bibinfo{author}{Sch\"afer, M.}, \bibinfo{author}{Barnea, N.} \& \bibinfo{author}{Gal, A.}
\newblock \bibinfo{journal}{\bibinfo{title}{In-medium $\mathrm{\ensuremath{\Lambda}}$ isospin impurity from charge symmetry breaking in the ${}_{\mathrm{\ensuremath{\Lambda}}}^{4}\mathrm{H}\text{\ensuremath{-}}{}_{\mathrm{\ensuremath{\Lambda}}}^{4}\mathrm{He}$ mirror hypernuclei}}.
\newblock {\emph{\JournalTitle{Phys. Rev. C}}} \textbf{\bibinfo{volume}{106}}, \bibinfo{pages}{L031001}, \doiprefix\url{10.1103/PhysRevC.106.L031001} (\bibinfo{year}{2022}).

\bibitem{Contessi2019csf}
\bibinfo{author}{Contessi, L.}, \bibinfo{author}{Sch\"afer, M.}, \bibinfo{author}{Barnea, N.}, \bibinfo{author}{Gal, A.} \& \bibinfo{author}{Mare\v{s}, J.}
\newblock \bibinfo{journal}{\bibinfo{title}{{The onset of $\Lambda\Lambda$-hypernuclear binding}}}.
\newblock {\emph{\JournalTitle{Phys. Lett. B}}} \textbf{\bibinfo{volume}{797}}, \bibinfo{pages}{134893}, \doiprefix\url{10.1016/j.physletb.2019.134893} (\bibinfo{year}{2019}).
\newblock \eprint{1905.06775}.

\bibitem{Schafer2020rba}
\bibinfo{author}{Schafer, M.}, \bibinfo{author}{Bazak, B.}, \bibinfo{author}{Barnea, N.} \& \bibinfo{author}{Mares, J.}
\newblock \bibinfo{journal}{\bibinfo{title}{{Nature of the $\Lambda nn$ $(J^\pi=1/2^+, I=1)$ and ${\rm ^3_\Lambda H^*} (J^\pi=3/2^+, I=0)$ states}}}.
\newblock {\emph{\JournalTitle{Phys. Rev. C}}} \textbf{\bibinfo{volume}{103}}, \bibinfo{pages}{025204}, \doiprefix\url{10.1103/PhysRevC.103.025204} (\bibinfo{year}{2021}).
\newblock \eprint{2007.10264}.

\bibitem{Schfer2021ConsequencesOI}
\bibinfo{author}{Schafer, M.}, \bibinfo{author}{Bazak, B.}, \bibinfo{author}{Barnea, N.}, \bibinfo{author}{Gal, A.} \& \bibinfo{author}{Mares, J.}
\newblock \bibinfo{journal}{\bibinfo{title}{Consequences of increased hypertriton binding for s-shell $\lambda$-hypernuclear systems}}.
\newblock {\emph{\JournalTitle{Physical Review C}}}  (\bibinfo{year}{2021}).

\bibitem{SVMPRL}
\bibinfo{author}{Contessi, L.}, \bibinfo{author}{Barnea, N.} \& \bibinfo{author}{Gal, A.}
\newblock \bibinfo{journal}{\bibinfo{title}{Resolving the $\mathrm{\ensuremath{\Lambda}}$ hypernuclear overbinding problem in pionless effective field theory}}.
\newblock {\emph{\JournalTitle{Phys. Rev. Lett.}}} \textbf{\bibinfo{volume}{121}}, \bibinfo{pages}{102502}, \doiprefix\url{10.1103/PhysRevLett.121.102502} (\bibinfo{year}{2018}).

\bibitem{neuralsci}
\bibinfo{author}{Carleo, G.} \& \bibinfo{author}{Troyer, M.}
\newblock \bibinfo{journal}{\bibinfo{title}{Solving the quantum many-body problem with artificial neural networks}}.
\newblock {\emph{\JournalTitle{Science}}} \textbf{\bibinfo{volume}{355}}, \bibinfo{pages}{602--606}, \doiprefix\url{10.1126/science.aag2302} (\bibinfo{year}{2017}).
\newblock \eprint{https://www.science.org/doi/pdf/10.1126/science.aag2302}.

\bibitem{PhysRevB.61.2599}
\bibinfo{author}{Sorella, S.} \& \bibinfo{author}{Capriotti, L.}
\newblock \bibinfo{journal}{\bibinfo{title}{Green function monte carlo with stochastic reconfiguration: An effective remedy for the sign problem}}.
\newblock {\emph{\JournalTitle{Phys. Rev. B}}} \textbf{\bibinfo{volume}{61}}, \bibinfo{pages}{2599--2612}, \doiprefix\url{10.1103/PhysRevB.61.2599} (\bibinfo{year}{2000}).

\bibitem{Stokes2020}
\bibinfo{author}{Stokes, J.}, \bibinfo{author}{Izaac, J.}, \bibinfo{author}{Killoran, N.} \& \bibinfo{author}{Carleo, G.}
\newblock \bibinfo{journal}{\bibinfo{title}{Quantum {N}atural {G}radient}}.
\newblock {\emph{\JournalTitle{{Quantum}}}} \textbf{\bibinfo{volume}{4}}, \bibinfo{pages}{269}, \doiprefix\url{10.22331/q-2020-05-25-269} (\bibinfo{year}{2020}).

\bibitem{deepwf}
\bibinfo{author}{Han, J.}, \bibinfo{author}{Zhang, L.} \& \bibinfo{author}{E, W.}
\newblock \bibinfo{journal}{\bibinfo{title}{Solving many-electron schrödinger equation using deep neural networks}}.
\newblock {\emph{\JournalTitle{Journal of Computational Physics}}} \textbf{\bibinfo{volume}{399}}, \bibinfo{pages}{108929}, \doiprefix\url{https://doi.org/10.1016/j.jcp.2019.108929} (\bibinfo{year}{2019}).

\bibitem{Ferminet}
\bibinfo{author}{Pfau, D.}, \bibinfo{author}{Spencer, J.~S.}, \bibinfo{author}{Matthews, A. G. D.~G.} \& \bibinfo{author}{Foulkes, W. M.~C.}
\newblock \bibinfo{journal}{\bibinfo{title}{Ab initio solution of the many-electron schr\"odinger equation with deep neural networks}}.
\newblock {\emph{\JournalTitle{Phys. Rev. Res.}}} \textbf{\bibinfo{volume}{2}}, \bibinfo{pages}{033429}, \doiprefix\url{10.1103/PhysRevResearch.2.033429} (\bibinfo{year}{2020}).

\bibitem{vonglehn2023selfattentionansatzabinitioquantum}
\bibinfo{author}{von Glehn, I.}, \bibinfo{author}{Spencer, J.~S.} \& \bibinfo{author}{Pfau, D.}
\newblock \bibinfo{title}{A self-attention ansatz for ab-initio quantum chemistry} (\bibinfo{year}{2023}).
\newblock \eprint{2211.13672}.

\bibitem{PauliNet}
\bibinfo{author}{Hermann, J.}, \bibinfo{author}{Schätzle, Z.} \& \bibinfo{author}{Noé, F.}
\newblock \bibinfo{journal}{\bibinfo{title}{Deep-neural-network solution of the electronic {Schrödinger} equation}}.
\newblock {\emph{\JournalTitle{Nature Chemistry}}} \textbf{\bibinfo{volume}{12}}, \bibinfo{pages}{891--897}, \doiprefix\url{10.1038/s41557-020-0544-y} (\bibinfo{year}{2020}).

\bibitem{Symmetries}
\bibinfo{author}{Li, Z.} \emph{et~al.}
\newblock \bibinfo{journal}{\bibinfo{title}{Spin-symmetry-enforced solution of the many-body {Schrödinger} equation with a deep neural network}}.
\newblock {\emph{\JournalTitle{Nature Computational Science}}} \textbf{\bibinfo{volume}{4}}, \bibinfo{pages}{910--919}, \doiprefix\url{10.1038/s43588-024-00730-4} (\bibinfo{year}{2024}).

\bibitem{Laplacian}
\bibinfo{author}{Li, R.} \emph{et~al.}
\newblock \bibinfo{journal}{\bibinfo{title}{A computational framework for neural network-based variational {Monte Carlo} with {Forward Laplacian}}}.
\newblock {\emph{\JournalTitle{Nature Machine Intelligence}}} \textbf{\bibinfo{volume}{6}}, \bibinfo{pages}{209--219}, \doiprefix\url{10.1038/s42256-024-00794-x} (\bibinfo{year}{2024}).

\bibitem{FermiSci}
\bibinfo{author}{Pfau, D.}, \bibinfo{author}{Axelrod, S.}, \bibinfo{author}{Sutterud, H.}, \bibinfo{author}{von Glehn, I.} \& \bibinfo{author}{Spencer, J.~S.}
\newblock \bibinfo{journal}{\bibinfo{title}{Accurate computation of quantum excited states with neural networks}}.
\newblock {\emph{\JournalTitle{Science}}} \textbf{\bibinfo{volume}{385}}, \bibinfo{pages}{eadn0137}, \doiprefix\url{10.1126/science.adn0137} (\bibinfo{year}{2024}).

\bibitem{chen_empowering_2024}
\bibinfo{author}{Chen, A.} \& \bibinfo{author}{Heyl, M.}
\newblock \bibinfo{journal}{\bibinfo{title}{Empowering deep neural quantum states through efficient optimization}}.
\newblock {\emph{\JournalTitle{Nature Physics}}} \bibinfo{pages}{1--6}, \doiprefix\url{10.1038/s41567-024-02566-1} (\bibinfo{year}{2024}).
\newblock \bibinfo{note}{Publisher: Nature Publishing Group}.

\bibitem{luo_simulating_2024}
\bibinfo{author}{Luo, D.}, \bibinfo{author}{Dai, D.~D.} \& \bibinfo{author}{Fu, L.}
\newblock \bibinfo{title}{Simulating moiré quantum matter with neural network}, \doiprefix\url{10.48550/arXiv.2406.17645} (\bibinfo{year}{2024}).
\newblock \bibinfo{note}{ArXiv:2406.17645}.

\bibitem{teng_solving_2025}
\bibinfo{author}{Teng, Y.}, \bibinfo{author}{Dai, D.~D.} \& \bibinfo{author}{Fu, L.}
\newblock \bibinfo{journal}{\bibinfo{title}{Solving the fractional quantum {Hall} problem with self-attention neural network}}.
\newblock {\emph{\JournalTitle{Physical Review B}}} \textbf{\bibinfo{volume}{111}}, \bibinfo{pages}{205117}, \doiprefix\url{10.1103/PhysRevB.111.205117} (\bibinfo{year}{2025}).

\bibitem{linteau_phase_2025}
\bibinfo{author}{Linteau, D.}, \bibinfo{author}{Pescia, G.}, \bibinfo{author}{Nys, J.}, \bibinfo{author}{Carleo, G.} \& \bibinfo{author}{Holzmann, M.}
\newblock \bibinfo{journal}{\bibinfo{title}{Phase {Diagram} and {Crystal} {Melting} of {Helium}-4 in {Two} {Dimensions}}}.
\newblock {\emph{\JournalTitle{Physical Review Letters}}} \textbf{\bibinfo{volume}{134}}, \bibinfo{pages}{246001}, \doiprefix\url{10.1103/v1g7-m9k4} (\bibinfo{year}{2025}).

\bibitem{qian_describing_2025}
\bibinfo{author}{Qian, Y.} \emph{et~al.}
\newblock \bibinfo{journal}{\bibinfo{title}{Describing {Landau} {Level} {Mixing} in {Fractional} {Quantum} {Hall} {States} with {Deep} {Learning}}}.
\newblock {\emph{\JournalTitle{Physical Review Letters}}} \textbf{\bibinfo{volume}{134}}, \bibinfo{pages}{176503}, \doiprefix\url{10.1103/PhysRevLett.134.176503} (\bibinfo{year}{2025}).

\bibitem{li_emergent_2024}
\bibinfo{author}{Li, X.}, \bibinfo{author}{Qian, Y.}, \bibinfo{author}{Ren, W.}, \bibinfo{author}{Xu, Y.} \& \bibinfo{author}{Chen, J.}
\newblock \bibinfo{title}{Emergent {Wigner} phases in moiré superlattice from deep learning}, \doiprefix\url{10.48550/arXiv.2406.11134} (\bibinfo{year}{2024}).
\newblock \bibinfo{note}{ArXiv:2406.11134 [physics]}.

\bibitem{li_deep_2025}
\bibinfo{author}{Li, X.} \emph{et~al.}
\newblock \bibinfo{title}{Deep {Learning} {Sheds} {Light} on {Integer} and {Fractional} {Topological} {Insulators}}, \doiprefix\url{10.48550/arXiv.2503.11756} (\bibinfo{year}{2025}).
\newblock \bibinfo{note}{ArXiv:2503.11756 [cond-mat]}.

\bibitem{linteau_universal_2025}
\bibinfo{author}{Linteau, D.}, \bibinfo{author}{Moroni, S.}, \bibinfo{author}{Carleo, G.} \& \bibinfo{author}{Holzmann, M.}
\newblock \bibinfo{title}{Universal neural wave functions for high-pressure hydrogen}, \doiprefix\url{10.48550/arXiv.2504.07062} (\bibinfo{year}{2025}).
\newblock \bibinfo{note}{ArXiv:2504.07062 [cond-mat]}.

\bibitem{v2007}
\bibinfo{author}{Adams, C.}, \bibinfo{author}{Carleo, G.}, \bibinfo{author}{Lovato, A.} \& \bibinfo{author}{Rocco, N.}
\newblock \bibinfo{journal}{\bibinfo{title}{Variational monte carlo calculations of ${A}\ensuremath{\le}4$ nuclei with an artificial neural-network correlator ansatz}}.
\newblock {\emph{\JournalTitle{Phys. Rev. Lett.}}} \textbf{\bibinfo{volume}{127}}, \bibinfo{pages}{022502}, \doiprefix\url{10.1103/PhysRevLett.127.022502} (\bibinfo{year}{2021}).

\bibitem{A6}
\bibinfo{author}{Gnech, A.} \emph{et~al.}
\newblock \bibinfo{journal}{\bibinfo{title}{Nuclei with up to a=6 nucleons with artificial neural network wave functions}}.
\newblock {\emph{\JournalTitle{Few-Body Systems}}} \textbf{\bibinfo{volume}{63}}, \bibinfo{pages}{7}, \doiprefix\url{10.1007/s00601-021-01706-0} (\bibinfo{year}{2021}).

\bibitem{Hidden}
\bibinfo{author}{Lovato, A.}, \bibinfo{author}{Adams, C.}, \bibinfo{author}{Carleo, G.} \& \bibinfo{author}{Rocco, N.}
\newblock \bibinfo{journal}{\bibinfo{title}{Hidden-nucleons neural-network quantum states for the nuclear many-body problem}}.
\newblock {\emph{\JournalTitle{Phys. Rev. Res.}}} \textbf{\bibinfo{volume}{4}}, \bibinfo{pages}{043178}, \doiprefix\url{10.1103/PhysRevResearch.4.043178} (\bibinfo{year}{2022}).

\bibitem{Feynmannet}
\bibinfo{author}{Yang, Y.~L.} \& \bibinfo{author}{Zhao, P.~W.}
\newblock \bibinfo{journal}{\bibinfo{title}{Deep-neural-network approach to solving the ab initio nuclear structure problem}}.
\newblock {\emph{\JournalTitle{Phys. Rev. C}}} \textbf{\bibinfo{volume}{107}}, \bibinfo{pages}{034320}, \doiprefix\url{10.1103/PhysRevC.107.034320} (\bibinfo{year}{2023}).

\bibitem{distill}
\bibinfo{author}{Gnech, A.}, \bibinfo{author}{Fore, B.}, \bibinfo{author}{Tropiano, A.~J.} \& \bibinfo{author}{Lovato, A.}
\newblock \bibinfo{journal}{\bibinfo{title}{Distilling the essential elements of nuclear binding via neural-network quantum states}}.
\newblock {\emph{\JournalTitle{Phys. Rev. Lett.}}} \textbf{\bibinfo{volume}{133}}, \bibinfo{pages}{142501}, \doiprefix\url{10.1103/PhysRevLett.133.142501} (\bibinfo{year}{2024}).

\bibitem{Kim2024}
\bibinfo{author}{Kim, J.} \emph{et~al.}
\newblock \bibinfo{journal}{\bibinfo{title}{Neural-network quantum states for ultra-cold fermi gases}}.
\newblock {\emph{\JournalTitle{Communications Physics}}} \textbf{\bibinfo{volume}{7}}, \bibinfo{pages}{148}, \doiprefix\url{10.1038/s42005-024-01613-w} (\bibinfo{year}{2024}).

\bibitem{PhysRevX.14.021030}
\bibinfo{author}{Lou, W.~T.} \emph{et~al.}
\newblock \bibinfo{journal}{\bibinfo{title}{Neural wave functions for superfluids}}.
\newblock {\emph{\JournalTitle{Phys. Rev. X}}} \textbf{\bibinfo{volume}{14}}, \bibinfo{pages}{021030}, \doiprefix\url{10.1103/PhysRevX.14.021030} (\bibinfo{year}{2024}).

\bibitem{elegas}
\bibinfo{author}{Pescia, G.}, \bibinfo{author}{Nys, J.}, \bibinfo{author}{Kim, J.}, \bibinfo{author}{Lovato, A.} \& \bibinfo{author}{Carleo, G.}
\newblock \bibinfo{journal}{\bibinfo{title}{Message-passing neural quantum states for the homogeneous electron gas}}.
\newblock {\emph{\JournalTitle{Phys. Rev. B}}} \textbf{\bibinfo{volume}{110}}, \bibinfo{pages}{035108}, \doiprefix\url{10.1103/PhysRevB.110.035108} (\bibinfo{year}{2024}).

\bibitem{crust}
\bibinfo{author}{Fore, B.}, \bibinfo{author}{Kim, J.}, \bibinfo{author}{Hjorth-Jensen, M.} \& \bibinfo{author}{Lovato, A.}
\newblock \bibinfo{title}{Investigating the crust of neutron stars with neural-network quantum states} (\bibinfo{year}{2024}).
\newblock \eprint{2407.21207}.

\bibitem{PhysRevLett.122.226401}
\bibinfo{author}{Luo, D.} \& \bibinfo{author}{Clark, B.~K.}
\newblock \bibinfo{journal}{\bibinfo{title}{Backflow transformations via neural networks for quantum many-body wave functions}}.
\newblock {\emph{\JournalTitle{Phys. Rev. Lett.}}} \textbf{\bibinfo{volume}{122}}, \bibinfo{pages}{226401}, \doiprefix\url{10.1103/PhysRevLett.122.226401} (\bibinfo{year}{2019}).

\bibitem{MPNN}
\bibinfo{author}{Gilmer, J.}, \bibinfo{author}{Schoenholz, S.~S.}, \bibinfo{author}{Riley, P.~F.}, \bibinfo{author}{Vinyals, O.} \& \bibinfo{author}{Dahl, G.~E.}
\newblock \bibinfo{title}{Neural message passing for quantum chemistry} (\bibinfo{year}{2017}).
\newblock \eprint{1704.01212}.

\bibitem{Entwistle2023}
\bibinfo{author}{Entwistle, M.~T.}, \bibinfo{author}{Schätzle, Z.}, \bibinfo{author}{Erdman, P.~A.}, \bibinfo{author}{Hermann, J.} \& \bibinfo{author}{Noé, F.}
\newblock \bibinfo{journal}{\bibinfo{title}{Electronic excited states in deep variational monte carlo}}.
\newblock {\emph{\JournalTitle{Nature Communications}}} \textbf{\bibinfo{volume}{14}}, \bibinfo{pages}{274}, \doiprefix\url{10.1038/s41467-022-35534-5} (\bibinfo{year}{2023}).

\bibitem{pionless}
\bibinfo{author}{Schiavilla, R.} \emph{et~al.}
\newblock \bibinfo{journal}{\bibinfo{title}{Two- and three-nucleon contact interactions and ground-state energies of light- and medium-mass nuclei}}.
\newblock {\emph{\JournalTitle{Phys. Rev. C}}} \textbf{\bibinfo{volume}{103}}, \bibinfo{pages}{054003}, \doiprefix\url{10.1103/PhysRevC.103.054003} (\bibinfo{year}{2021}).

\bibitem{thomas}
\bibinfo{author}{Thomas, L.~H.}
\newblock \bibinfo{journal}{\bibinfo{title}{The interaction between a neutron and a proton and the structure of ${\mathrm{h}}^{3}$}}.
\newblock {\emph{\JournalTitle{Phys. Rev.}}} \textbf{\bibinfo{volume}{47}}, \bibinfo{pages}{903--909}, \doiprefix\url{10.1103/PhysRev.47.903} (\bibinfo{year}{1935}).

\bibitem{eckert2021hypernuclides}
\bibinfo{author}{Eckert, P.}, \bibinfo{author}{Achenbach, P.} \emph{et~al.}
\newblock \bibinfo{title}{Chart of hypernuclides --- hypernuclear structure and decay data}.
\newblock \bibinfo{howpublished}{\url{https://hypernuclei.kph.uni-mainz.de}} (\bibinfo{year}{2021}).

\bibitem{didonna2025hypernucleineuralnetworkquantum}
\bibinfo{author}{Donna, A.~D.}, \bibinfo{author}{Contessi, L.}, \bibinfo{author}{Lovato, A.} \& \bibinfo{author}{Pederiva, F.}
\newblock \bibinfo{title}{Hypernuclei with neural network quantum states} (\bibinfo{year}{2025}).
\newblock \eprint{2507.16994}.

\bibitem{Cobis_1997}
\bibinfo{author}{Cobis, A.}, \bibinfo{author}{Jensen, A.~S.} \& \bibinfo{author}{Fedorov, D.~V.}
\newblock \bibinfo{journal}{\bibinfo{title}{The simplest strange three-body halo}}.
\newblock {\emph{\JournalTitle{Journal of Physics G: Nuclear and Particle Physics}}} \textbf{\bibinfo{volume}{23}}, \bibinfo{pages}{401}, \doiprefix\url{10.1088/0954-3899/23/4/002} (\bibinfo{year}{1997}).

\bibitem{PhysRevC.59.2351}
\bibinfo{author}{Hiyama, E.}, \bibinfo{author}{Kamimura, M.}, \bibinfo{author}{Miyazaki, K.} \& \bibinfo{author}{Motoba, T.}
\newblock \bibinfo{journal}{\bibinfo{title}{\ensuremath{\gamma} transitions in ${A}=7$ hypernuclei and a possible derivation of hypernuclear size}}.
\newblock {\emph{\JournalTitle{Phys. Rev. C}}} \textbf{\bibinfo{volume}{59}}, \bibinfo{pages}{2351--2360}, \doiprefix\url{10.1103/PhysRevC.59.2351} (\bibinfo{year}{1999}).

\bibitem{PhysRevLett.134.162503}
\bibinfo{author}{Shen, S.}, \bibinfo{author}{Elhatisari, S.}, \bibinfo{author}{Lee, D.}, \bibinfo{author}{Mei\ss{}ner, U.-G.} \& \bibinfo{author}{Ren, Z.}
\newblock \bibinfo{journal}{\bibinfo{title}{Ab initio study of the beryllium isotopes $^{7}\mathrm{Be}$ to $^{12}\mathrm{Be}$}}.
\newblock {\emph{\JournalTitle{Phys. Rev. Lett.}}} \textbf{\bibinfo{volume}{134}}, \bibinfo{pages}{162503}, \doiprefix\url{10.1103/PhysRevLett.134.162503} (\bibinfo{year}{2025}).

\bibitem{PhysRevLett.89.182501}
\bibinfo{author}{Wiringa, R.~B.} \& \bibinfo{author}{Pieper, S.~C.}
\newblock \bibinfo{journal}{\bibinfo{title}{Evolution of nuclear spectra with nuclear forces}}.
\newblock {\emph{\JournalTitle{Phys. Rev. Lett.}}} \textbf{\bibinfo{volume}{89}}, \bibinfo{pages}{182501}, \doiprefix\url{10.1103/PhysRevLett.89.182501} (\bibinfo{year}{2002}).

\bibitem{niu2025signproblemfreenuclearquantummonte}
\bibinfo{author}{Niu, Z.-W.} \& \bibinfo{author}{Lu, B.-N.}
\newblock \bibinfo{title}{Sign-problem-free nuclear quantum monte carlo} (\bibinfo{year}{2025}).
\newblock \eprint{2506.12874}.

\bibitem{IPATOV200960}
\bibinfo{author}{Ipatov, A.}, \bibinfo{author}{Cordova, F.}, \bibinfo{author}{Doriol, L.~J.} \& \bibinfo{author}{Casida, M.~E.}
\newblock \bibinfo{journal}{\bibinfo{title}{Excited-state spin-contamination in time-dependent density-functional theory for molecules with open-shell ground states}}.
\newblock {\emph{\JournalTitle{Journal of Molecular Structure: THEOCHEM}}} \textbf{\bibinfo{volume}{914}}, \bibinfo{pages}{60--73}, \doiprefix\url{https://doi.org/10.1016/j.theochem.2009.07.036} (\bibinfo{year}{2009}).
\newblock \bibinfo{note}{Time-dependent density-functional theory for molecules and molecular solids}.

\bibitem{Li2024}
\bibinfo{author}{Li, Z.} \emph{et~al.}
\newblock \bibinfo{journal}{\bibinfo{title}{Spin-symmetry-enforced solution of the many-body schrödinger equation with a deep neural network}}.
\newblock {\emph{\JournalTitle{Nature Computational Science}}} \textbf{\bibinfo{volume}{4}}, \bibinfo{pages}{910--919}, \doiprefix\url{10.1038/s43588-024-00730-4} (\bibinfo{year}{2024}).

\bibitem{schnet}
\bibinfo{author}{Schütt, K.~T.} \emph{et~al.}
\newblock \bibinfo{title}{Schnet: A continuous-filter convolutional neural network for modeling quantum interactions} (\bibinfo{year}{2017}).
\newblock \eprint{1706.08566}.

\bibitem{resnet}
\bibinfo{author}{He, K.}, \bibinfo{author}{Zhang, X.}, \bibinfo{author}{Ren, S.} \& \bibinfo{author}{Sun, J.}
\newblock \bibinfo{title}{Deep residual learning for image recognition} (\bibinfo{year}{2015}).
\newblock \eprint{1512.03385}.

\bibitem{Lu:2018bat}
\bibinfo{author}{Lu, B.-N.} \emph{et~al.}
\newblock \bibinfo{journal}{\bibinfo{title}{{Essential elements for nuclear binding}}}.
\newblock {\emph{\JournalTitle{Phys. Lett. B}}} \textbf{\bibinfo{volume}{797}}, \bibinfo{pages}{134863}, \doiprefix\url{10.1016/j.physletb.2019.134863} (\bibinfo{year}{2019}).
\newblock \eprint{1812.10928}.

\end{thebibliography}

\clearpage
\setcounter{section}{0}
\setcounter{figure}{0}
\setcounter{enumiv}{0}

\twocolumn[
\begin{@twocolumnfalse}
\begin{center}
	\noindent\LARGE\textbf{Supplementary Material} \\ % 使用\LARGE加大字体
\end{center}
This supplementary material provides further details for this article. The numerical results and original data are presented in the form of detailed tables and combined figures.
	\bigskip
\section{Convergence pattern with Hamiltonian \enquote{o}}
    \begin{center}
	\centerline{\includegraphics[width=0.72\textwidth]{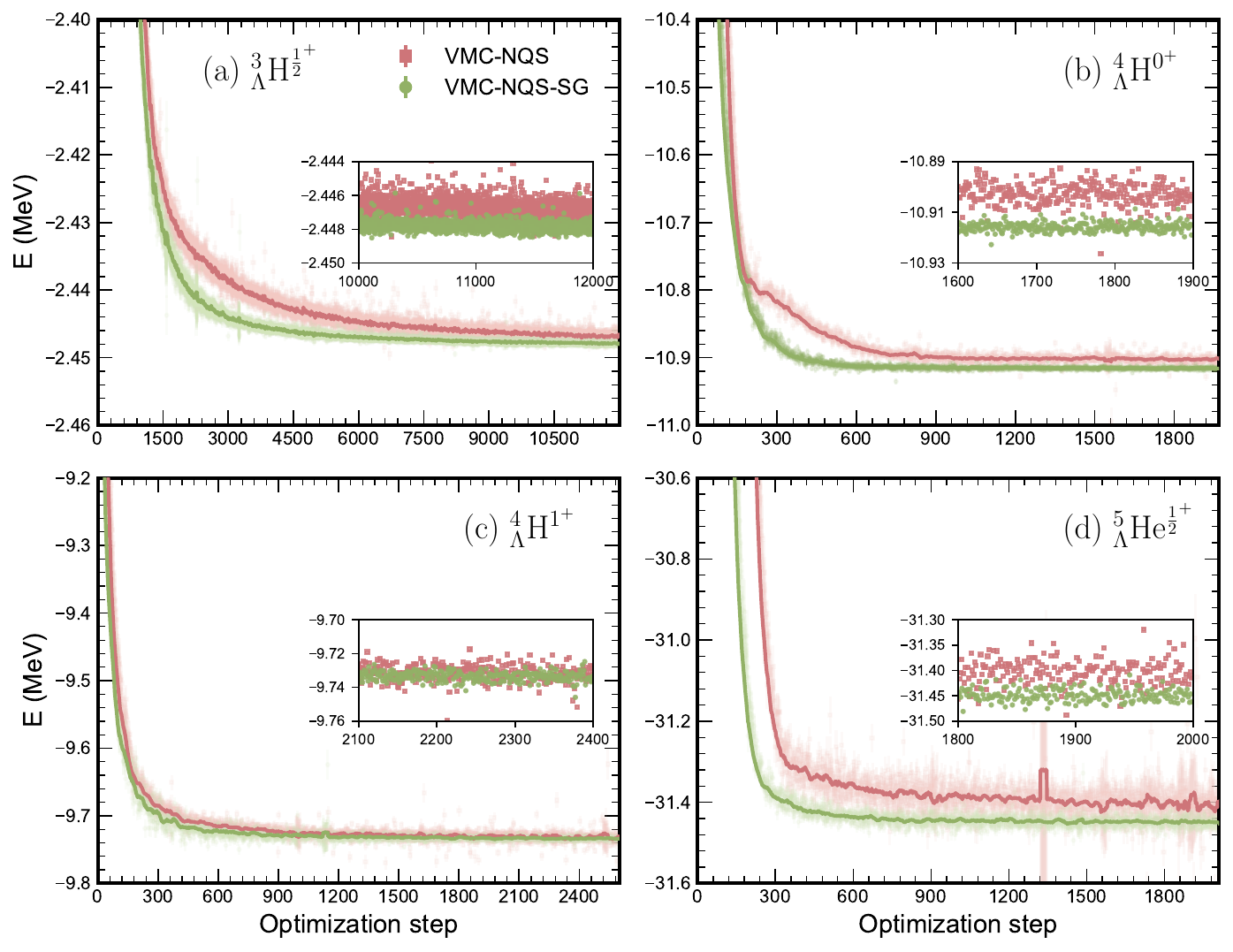}}
	\captionof{figure}{FIG. 1. Energy convergence patterns of (a) $^3_\Lambda$H$^{\frac{1}{2}^+}$, (b) $^4_\Lambda$H$^{0^+}$, (c) $^4_\Lambda$H$^{1^+}$ and  (d) $^5_\Lambda$He$^{\frac{1}{2}^+}$, obtained with Hmiltonian \enquote{o}. Raw data points from VMC-NQS are shown as red solid squares, while green solid circles denote that with spinor grouping technique (VMC-NQS-SG).  The corresponding sliding curve and error bars are obtained by calculating the mean and standard deviation over a 25-step sliding window.
    Inset zoomed regions highlight late-stage convergence behavior.}
	\label{fig:subconvergence}-
    \end{center}
In Fig. \ref{fig:subconvergence}, we provide the energy convergence pattern for $^3_\Lambda$H$^{\frac{1}{2}^+}$,  $^4_\Lambda$H$^{0^+}$,  $^4_\Lambda$H$^{1^+}$ and   $^5_\Lambda$He$^{\frac{1}{2}^+}$, calculated with model \enquote{o}. SG method significantly boosts both the accuracy and precision of VMC-NQS method.

\section{Numerical results of $s$-shell hypernuclei}
	
	\begin{center}
 % 手动添加标题
    \captionof{table}{Table 1. The $B_\Lambda$ and $r_{\Lambda N}$ values results from FeynmanNet, VMC-NQS and VMC-NQS-SG methods,  using Hamiltonian \enquote{s}. The results and errors  are taken as the mean and standard deviation over the last 100 iterations. $B_\Lambda$ results from SVM method \cite{PhysRevC.106.L031001} are provided as baseline. The experimental results are taken from Ref.~\cite{eckert2021hypernuclides}, with the four-body systems averaged.}
	\begin{tabular}{l c c c c c c c c}
		\toprule
		\text{Nucleus}                      & \multicolumn{5}{c}{\text{$B_\Lambda$ (MeV)}} & \multicolumn{3}{c}{\text{$r_{\Lambda N}$ (fm)}} \\
		\cmidrule(lr){2-6} \cmidrule(lr){7-9} & \text{FeynmanNet}&\text{VMC-NQS} & \text{VMC-NQS-SG} & SVM \cite{PhysRevC.106.L031001} &\text{Exp. \cite{eckert2021hypernuclides}}  &\text{FeynmanNet}  & \text{VMC-NQS} & \text{VMC-NQS-SG}                \\ \midrule
		$^3_\Lambda$H$^{\frac{1}{2}^+}$                  &  0.138(3)             & 0.1386(11)              & 0.1377(3)         & 0.13             & 0.164(43)     &8.734(9)        & 9.967(46)      &       9.414(39)           \\
		$^4_\Lambda$H$^{0^+}$                          & 2.266(9)      & 2.275(18)              & 2.274(2)
		& 2.2735             & 2.258(55)          & 3.494(2)   & 3.463(11)            &3.458(9)            \\
		$^4_\Lambda$H$^{1^+}$                     &1.041(7)          & 1.0455(77)             & 1.0462(41)         & 1.0255            & 1.011(72)   & 4.594(4)          & 4.448(17)        &4.416(17)                \\ 
		$^5_\Lambda$He$^{\frac{1}{2}^+}$             &2.371(23)                   & 2.415(49)             & 2.457(11)         & 2.4187        & 3.102(30)    & 3.345(2)             & 3.321(10)          &3.341(11)            \\ \bottomrule
	\end{tabular}
\end{center}
\end{@twocolumnfalse}
]
\onecolumn
\begin{center}
    \captionof{table}{Table 2. The $B_\Lambda$ and $r_{\Lambda N}$ values results from VMC-NQS and VMC-NQS-SG methods,  using Hamiltonian \enquote{o}. The results and errors  are taken as the mean and standard deviation over the last 100 iterations. The experimental results are taken from Ref.~\cite{eckert2021hypernuclides}, with the four-body systems averaged.}
	\begin{tabular}{l c c c c c}
			\toprule
			\text{Nucleus}                      & \multicolumn{3}{c}{\text{$B_\Lambda$ (MeV)}} & \multicolumn{2}{c}{\text{$r_{\Lambda N}$ (fm)}} \\
			\cmidrule(lr){2-4} \cmidrule(lr){5-6} & \text{VMC-NQS} & \text{VMC-NQS-SG}  &\text{Exp. \cite{eckert2021hypernuclides}}    & \text{VMC-NQS} & \text{VMC-NQS-SG}                \\ \midrule
		$^3_\Lambda$H$^{\frac{1}{2}^+}$                                 & 0.2050(4)              & 0.2059(2)                     & 0.164(43)             & 8.255(28)      &      8.587(34)           \\
		$^4_\Lambda$H$^{0^+}$                                & 2.4321(36)              &2.4408(14)
		& 2.258(55)             & 3.793(19)            &3.513(10))            \\
		$^4_\Lambda$H$^{1^+}$                                & 1.2582(90)             & 1.2588(12)                    & 1.011(72)             & 4.259(11)        &4.336(13)                \\ 
		$^5_\Lambda$He$^{\frac{1}{2}^+}$                                & 3.234(68)             & 3.281(13)                  & 3.102(30)             & 3.100(9)          &3.117(9)           \\ \bottomrule
		\end{tabular}
		
\end{center}
\section{Performance of SG method in $p$-shell hypernuclei}
\begin{figure*}[h]% 使用 figure* 跨栏环境
\centering
% 第一行三张图
\begin{minipage}{0.32\linewidth}
  \includegraphics[width=\linewidth]{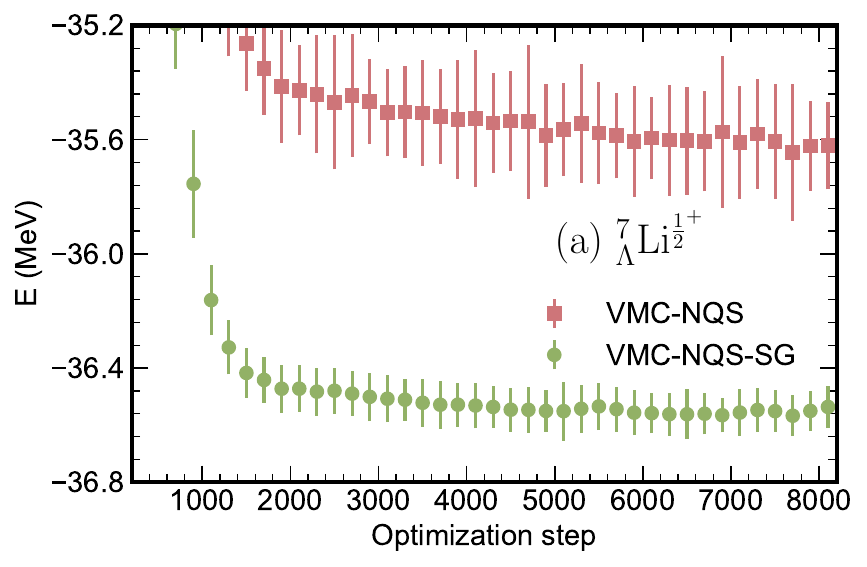}
\end{minipage}
\hfill
\begin{minipage}{0.32\linewidth}
  \includegraphics[width=\linewidth]{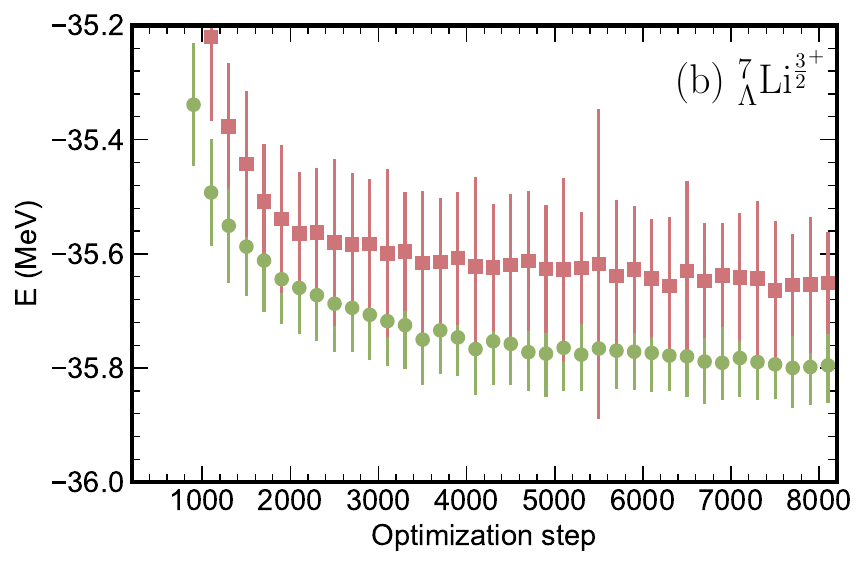}
\end{minipage}
\hfill
\begin{minipage}{0.32\linewidth}
  \includegraphics[width=\linewidth]{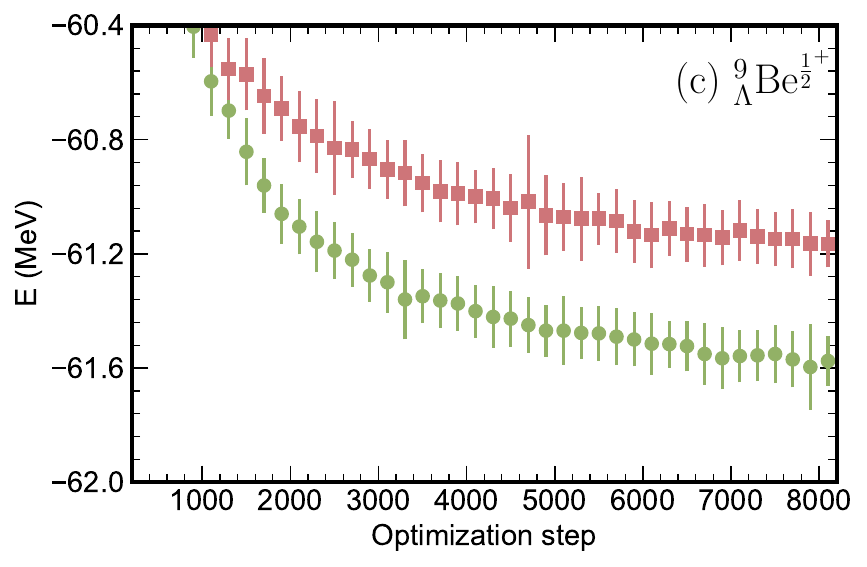}
\end{minipage}

\vspace{1em}

% 第二行三张图
\begin{minipage}{0.32\linewidth}
  \includegraphics[width=\linewidth]{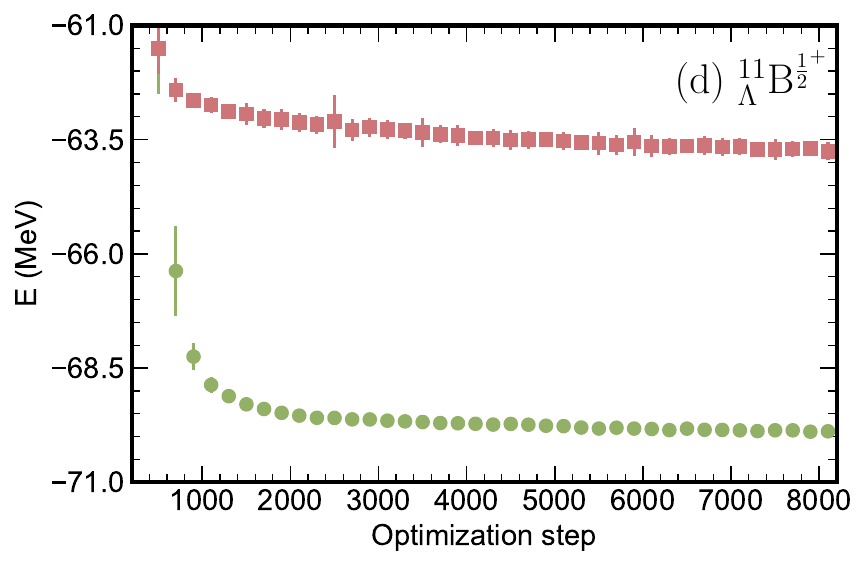}
\end{minipage}
\hfill
\begin{minipage}{0.32\linewidth}
  \includegraphics[width=\linewidth]{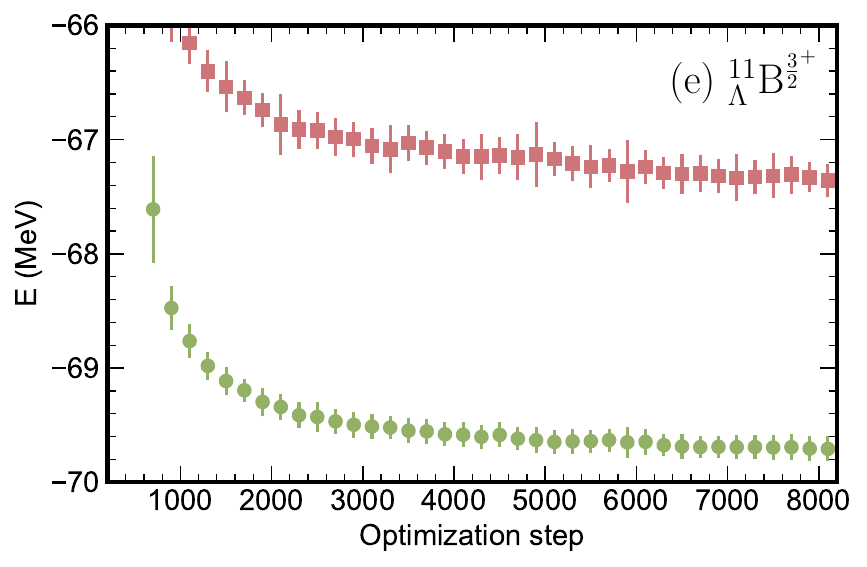}
\end{minipage}
\hfill
\begin{minipage}{0.32\linewidth}
  \includegraphics[width=\linewidth]{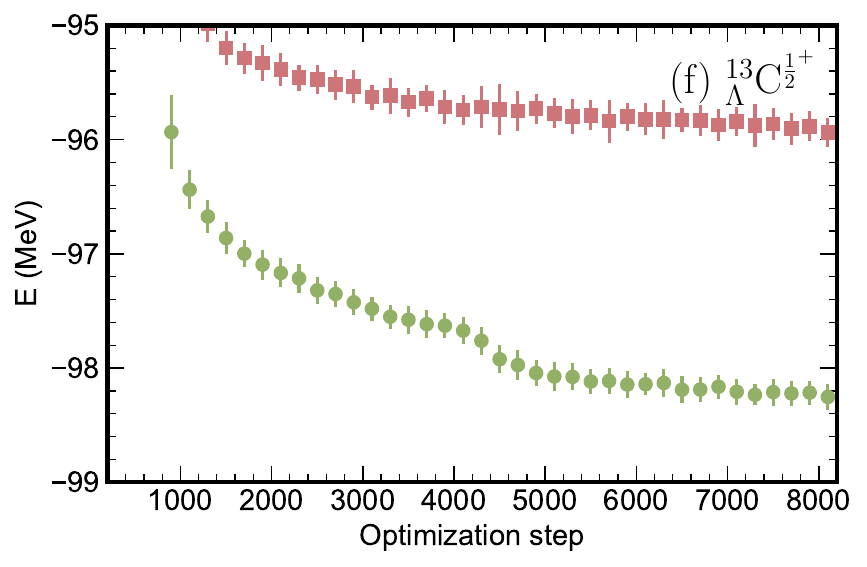}
\end{minipage}

\caption{FIG. 2.  Energy convergence patterns of (a) $^7_\Lambda$Li$^{\frac{1}{2}^+}$, (b) $^7_\Lambda$Li$^{\frac{3}{2}^+}$, (c) $^9_\Lambda$Be$^{\frac{1}{2}^+}$,  (d) $^{11}_\Lambda$B$^{\frac{1}{2}^+}$, (e) $^{11}_\Lambda$B$^{\frac{3}{2}^+}$ and (f) $^{13}_\Lambda$C$^{\frac{1}{2}^+}$ obtained with Hmiltonian \enquote{o}. Data points from VMC-NQS are shown as red solid squares, while green solid circles denote that with spinor grouping technique (VMC-NQS-SG).  For clarity, we bin the raw data points every 200 steps for all $p$-shell hypernuclei.  }
\label{fig:combined}
\end{figure*}

In Fig. \ref{fig:combined}, we highlight the growing efficacy of the SG method for heavier systems by comparing the energy convergence patterns of $p$-shell hypernuclei between the VMC-NQS and VMC-NQS-SG methods. All assessments are carried out under strictly identical conditions, including the same number of samples ($N=10,000$), number of determinants ($K=6$), learning rate and the MPNN architecture ($L=3$). Relative to the $s$-shell results (Fig. \ref{fig:subconvergence}), the SG method yields significantly larger improvements for $p$-shell hypernuclei over VMC-NQS. This is attributed to the significantly larger integration space required for calculating the energy expectation values for heavier nuclei. The SG method allows us to analytically integrate over the exponentially growing Hilbert space, rather than relying on Monte Carlo sampling. Furthermore, VMC-NQS-SG calculations require only the computation of block-diagonalized determinants, eliminating the need for full-matrix evaluation. Isospin sampling is also rendered unnecessary, thereby considerably reducing both memory consumption and computational complexity. The explicit labeling of isospins also facilitates the incorporation of mean-field information more conveniently. These features make SG method a promising tool for tackling heavier nuclei. 
\end{document}